\documentclass[trackchanges]{aastex62}
\usepackage[version=4]{mhchem}
\usepackage{float}
\usepackage{amsmath}
\usepackage[section]{placeins}

\graphicspath{{./}{figures/}}

\received{July 23rd, 2019}
\revised{September 4th, 2019}
\accepted{September 26th, 2019}
\submitjournal{ApJ}

\shorttitle{Low Temperature CH + CH$_{2}$O Kinetics}
\shortauthors{West et al.}
\begin{document}

%%%%%%%%%%%%%%%%%%%%%%%%%%%%%%%%%%%%%%
%%Title
%%%%%%%%%%%%%%%%%%%%%%%%%%%%%%%%%%%%%%
\title{
    Measurements of Low Temperature Rate Coefficients for the Reaction of CH with CH$_{2}$O and Application to Dark Cloud and AGB Stellar Wind Models.  
}

%%%%%%%%%%%%%%%%%%%%%%%%%%%%%%%%%%%%%%
%%Authors
%%%%%%%%%%%%%%%%%%%%%%%%%%%%%%%%%%%%%%
\correspondingauthor{Dwayne Heard}
\email{D.E.Heard@leeds.ac.uk}

\author[0000-0002-3847-8478]{Niclas A. West}
\affil{School of Chemistry, University of Leeds, Leeds, LS2 9JT, UK}

\author[0000-0001-5178-3656]{Tom J. Millar}
\affil{Astrophysics Research Centre, School of Mathematics and Physics, Queen’s University Belfast, University Road, Belfast BT7 1NN, UK}
\affil{Institute of Theory and Computation, Harvard-Smithsonian Center for Astrophysics, 60 Garden Street, Cambridge MA 02138, USA}

\author[0000-0001-9298-6265]{Marie Van de Sande}
\affil{Instituut voor Sterrenkunde, KU Leuven, Celestijnenlaan 200D, 3001 Leuven, Belgium}

\author{Edward Rutter}
\affil{School of Chemistry, University of Leeds, Leeds, LS2 9JT, UK}

\author[0000-0001-6710-4021]{Mark A. Blitz}
\affil{School of Chemistry, University of Leeds, Leeds, LS2 9JT, UK}

\author[0000-0002-5342-8612]{Leen Decin}
\affil{Instituut voor Sterrenkunde, KU Leuven, Celestijnenlaan 200D, 3001 Leuven, Belgium}

\author[0000-0002-0357-6238]{Dwayne E. Heard}
\affil{School of Chemistry, University of Leeds, Leeds, LS2 9JT, UK}

%%%%%%%%%%%%%%%%%%%%%%%%%%%%%%%%%%%%%%
%%Abstract
%%%%%%%%%%%%%%%%%%%%%%%%%%%%%%%%%%%%%%
\begin{abstract}
Rate coefficients have been measured for the reaction of CH radicals with formaldehyde, CH$_{2}$O, over the temperature range 31 - 133 K using a pulsed Laval nozzle apparatus combined with \textnormal{pulsed} laser photolysis and laser induced fluorescence spectroscopy.  
The rate coefficients are very large and display a distinct decrease with decreasing temperature below 70 K, although classical collision rate theory fails to reproduce this temperature dependence.  
The measured rate coefficients have been parameterized and used as input for astrochemical models for both dark cloud and AGB stellar outflow scenarios.  
The models predict a distinct change (up to a factor of two) in the abundance of ketene, H$_{2}$CCO, which is the major expected molecular product of the CH + CH$_{2}$O reaction.

\end{abstract}


%%%%%%%%%%%%%%%%%%%%%%%%%%%%%%%%%%%%%%
%%Introduction
%%%%%%%%%%%%%%%%%%%%%%%%%%%%%%%%%%%%%%
\section{Introduction} \label{sec:intro}

In order to adequately describe gas-phase astrochemistry within a model, it is necessary to have accurate knowledge of the rate coefficients for relevant species.  
Although progress has been made in the measurement of rate coefficients for neutral-neutral reactions at very low temperatures in the last 30 years since the invention of the CRESU technique\textnormal{ (French
acronym for “Cin\'{e}tique de R\'{e}action en Ecoulement Supersonique Uniforme”)}, there is still a limited database compared with those close to or above $\sim$298 K \citep{potapov2017uniform}.  
Theoretical prediction of rate coefficients at low temperatures is difficult owing to new reaction mechanisms sometimes becoming dominant at temperatures below $\sim$298 K. 
In addition, extrapolation of fits to experimental data above $\sim$298 K \citep{potapov2017uniform,heard2018rapid} can lead to errors  in predicted rate coefficients at low temperatures.  
For some systems, rate coefficients continue to increase with a decrease in temperature down to the lowest temperatures accessible when using the CRESU technique \citep{cooke2019experimental}.  
Indeed, the decrease in velocity with temperature often causes the importance of long-range molecular interactions to increase with a decrease in temperature.  
These mechanisms determine the temperature dependence of the  reaction rate coefficient until it reaches the rate, called the collision limit, at which every collision of the reagent gas molecules leads to reaction.  
For some reactions, the collision limit is reached at temperatures relevant to the modeling of astrochemical environments \citep{smith2006temperature}.  

One reaction with a rate coefficient near the collision limit and showing an inverse temperature dependence at $T >$ 298 K is that of methylidyne (CH, also called carbyne) with formaldehyde (CH$_{2}$O).  
Methylidyne was one of the first molecules detected in the interstellar medium.  
Its optical absorption lines (see \citet{mck41} for a discussion of its early history) were utilized to probe diffuse interstellar clouds while its radio wavelength transitions \citep{ryd73} confirmed its presence in dense molecular clouds.  
It is now known to be widespread with detections in interstellar shock waves and external galaxies \citep{danks84, sandell88, muller14}.  
Similarly, formaldehyde, CH$_2$O, the first discovered organic polyatomic molecule \citep{snyder1969microwave}, is also an ubiquitous interstellar molecule, seen in almost all types of interstellar molecular cloud as well as in the circumstellar envelopes (CSEs) of C-rich and O-rich Asymptotic Giant Branch (AGB) stars.  
In this paper, we study the fast reaction between CH and CH$_2$O, the major molecular product of which is ketene, H$_2$CCO \citep{nguyen2014theoretical}, and which is also a common molecule in dense interstellar clouds \citep{tur77, mat86, rui07}. 
Its formation can occur in the gas-phase via the radiative association CH$_3^+$ + CO $\longrightarrow$ H$_3$CCO$^+$ + h$\nu$ followed by dissociative recombination with electrons. 
The UMIST Rate12 database also includes the neutral-neutral reaction O + C$_2$H$_3$ $\longrightarrow$ H$_2$CCO + H with a large rate coefficient, 1.6 $\times$ 10$^{-10}$ cm$^3$ s$^{-1}$ \citep{tsa86}, although this value is uncertain and may be a factor of 3 too large \citep{bau05}. 
The KIDA database \citep{wak15} adopts an overall rate coefficient of \textnormal{1.1 $\times$} 10$^{-10}$ cm$^3$ s$^{-1}$ with channels to H$_2$CCO + H (\textnormal{55}\%), CH$_3$ + CO (\textnormal{36}\%) and C$_2$H$_2$ + OH (\textnormal{9}\%). 
In recent years, it has also been suggested that ketene formation is via surface chemistry on icy grains followed by desorption to the gas phase \citep{hud13, mai14}.

Furthermore, \textnormal{CH and CH$_{2}$O} have been observed in the atmospheres of Earth and Titan \citep{grosjean1993ambient,viskari2000seasonal,saxena2003production,krasnopolsky2009photochemical,nixon2010upper,atreya2010significance} as well as in combustion processes \citep{fenimore1971formation,carlier1986chemistry,miller1989mechanism,anderson1996sources,goulay2009cyclic}.  
Although the reagents CH and CH$_{2}$O as well as the major molecular products H$_{2}$CCO, CH$_{3}$, and CO have been observed in low temperature astrochemical environments, the rate coefficients for the reaction of CH + CH$_{2}$O have not been previously measured below 298 K. 

CH, in its ground electronic state, X$^{2}\Pi$, is highly reactive because it has both an unpaired electron and a lone pair of electrons.  
Additionally, the relatively large dipole moments of CH and CH$_{2}$O (given in Section \ref{sec:ExpResults}, Table \ref{tab:kcoll_parameters}) lead to stronger long-range interactions relevant to reaction mechanisms at low temperatures, making this an ideal system to study with the CRESU technique.  
Previously,  \cite{zabarnick1988temperature} measured an inverse temperature dependence of the rate coefficient, $k_{1}(T)$, for the reaction of CH + CH$_{2}$O at 298 K $<$ T $<$ 670 K by monitoring the rate of loss of CH via Laser-Induced Fluorescence (LIF).  
Subsequently, \cite{nguyen2014theoretical} modeled the reaction with Variational Transition State Theory (VTST) and Rice-Ramsperger-Kassel-Marcus (RRKM) Master Equation calculations and determined that it proceeds through a barrierless potential energy surface which causes the negative temperature dependence of the reaction rate coefficients above 300 K, similar to other barrierless reactions \citep{phillips1992rate}.  
The authors also concluded that the primary pathways of the reaction were:
\begin{subequations} %CH + CH2O Eqns
    \label{eqn:Rxn_CH_CH2O-Prod}
\begin{align} 
        \ce{&CH + CH2O ->[$k_{1a}(T)$] H + H2CCO} 
            \textrm{
                ($\sim$82\% at 300 K, $\Delta$H$^{\ominus}_{r}$ = -301.5 kJ mol}
            ^{-1}),
        \label{eqn:CH_CH2O-H_H2CCO}\\
        \ce{&CH + CH2O ->[$k_{1b}(T)$] CH3 + CO} 
            \textrm{ 
                ($\sim$16\% at 300 K, $\Delta$H$^{\ominus}_{r}$ = -442.8 kJ mol}
            ^{-1}),
        \label{eqn:CH_CH2O-CH3_CO}\\
        \ce{&CH + CH2O ->[$k_{1c}(T)$] $^{3}$CH2 + HCO} 
            \textrm{ 
                ($\lesssim$2\% at 300 K, $\Delta$H$^{\ominus}_{r}$ = -53.2 kJ mol}
            ^{-1}),
        \label{eqn:CH_CH2O-CH2_HCO}
\end{align}
\end{subequations}
\noindent where yields and enthalpies were calculated by \cite{nguyen2014theoretical} at low pressures. 
\cite{nguyen2014theoretical} also calculated that, above 300 K, the yield of channel \ref{eqn:CH_CH2O-H_H2CCO} \textnormal{decreased with increasing} temperature.  

Previous reactions of CH with alkenes and alkynes measured with the CRESU technique have exhibited rate coefficients with a negative temperature dependence above $T \sim$ 60 K, a maximum observed at $T \sim$ 60 K, and a positive $\sim T^{1/6}$ dependence for $T \lesssim$ 60 K \citep{canosa1997reactions,smith2006temperature}.  
This $\sim T^{1/6}$ dependence was attributed to the rate coefficient reaching the collision limit as determined by Classical Capture Theory (CCT).  
CCT improves upon the simple hard sphere collision model by using orientation-averaged, attractive, long-range potentials between molecules in order to approximate the extent to which the intermolecular potentials would deflect the molecules. 
Since second-order rate coefficients can be described as:
\begin{equation}
    k(T) = \sigma (T) \langle v(T) \rangle
\end{equation}

\noindent where $k(T)$ and $\sigma (T)$ are the temperature-dependent rate coefficient (cm$^{3}$s$^{-1}$) and the temperature-dependent cross section (cm$^{2}$), respectively, \textnormal{for a given process such as reaction, collision, quenching, etc}, and $\langle v(T) \rangle$ is the temperature-dependent average \textnormal{relative} velocity (cm s$^{-1}$), the hard sphere collision rate has a temperature dependence of $T^{1/2}$ due entirely to the temperature dependence of $\langle v(T) \rangle$ since the hard sphere $\sigma (T)$ is temperature independent.  
However, CCT predicts a different temperature dependence of the collision limit due to the dependence of $\sigma (T)$ on the form of the long-range attractive intermolecular potential.  
These potentials cause a larger deflection of molecules toward each other at low temperatures when average velocities of the molecules are slower, yielding a larger effective collision cross section.  
Unfortunately, the positive temperature dependence behavior of the reaction rate coefficients for CH with alkenes and alkynes measured by \cite{canosa1997reactions} and \cite{smith2006temperature} were not well defined due to the behavior only occurring at the lowest temperature accessible by the CRESU apparatuses \textnormal{in these experiments} ($\sim$23 K). 
Therefore, \textnormal{the data were only fitted with the standard modified Arrhenius equation} by \cite{canosa1997reactions} and \cite{smith2006temperature}; which has been previously expressed in two different forms:
\begin{subequations} %CH + CH2O Eqns
    \label{eqn:ModArrh}
\begin{align}
    k\left( T\right) &= 
    A\left( \dfrac 
        {T}
        {\tau}
    \right)^{n} exp\left( -\dfrac 
        {E_{a}}
        {RT}
    \right)
    \label{eqn:ModArrhClassic}\\
    k\left( T\right) &=\alpha \left( \dfrac 
        {T}
        {300}
    \right)^{\beta } exp\left( -\dfrac 
        {\gamma}
        {T}
    \right),
    \label{eqn:ModArrhAstro}
\end{align}
\end{subequations}

\noindent where $A$ is a pre-exponential factor (cm$^{3}$s$^{-1}$), $\tau$ is usually either defined as $\sim$300 K or 1 K, $n$ is a constant, $E_{a}$ is the activation energy, \textnormal{$R$ is the ideal gas constant}, and $\alpha$, $\beta$, and $\gamma$ are versions of the parameters often utilized in astrochemical modeling. 
Since $\tau =$ 300 K is \textnormal{employed} in the UMIST RATE12 and KIDA astrochemical databases, and $\tau =$ 1 K is often \textnormal{employed} in fits of low temperature rate coefficients, the two types of fits can be converted using:   
\begin{equation} %Conversion of Modified Arrhenius Equations
    A_{\tau = 300} = 
    300^{n}A_{\tau = 1}
    \label{eqn:Convert_ModArrh}
\end{equation}

\noindent Since it is possible to fit experimental temperature dependencies with modified Arrhenius equations which have very different forms to those predicted by collision rate models (for instance a $\approx T^{1/6}$ dependence predicted by capture theory), \cite{canosa1997reactions} stated that their modified Arrhenius fit should not be extrapolated below 23 K, the lowest temperature measured in their experiments.  

Several theoretical approaches have been developed to determine the temperature dependence of the collision limit.  
One method, CCT, predicts the rate coefficients of collisions controlled by the orientation-averaged long-range potential between two molecular species.  
Since the $r^{-6}$ intermolecular potentials due to dipole-dipole ($D-D$), dipole-induced-dipole ($D-iD$), and London dispersion ($Disp$) forces yield a good first-order approximation of the long-range potential between the many neutral species, the collision rate coefficient predicted by CCT, $k_{coll}(T)$, is:
\begin{equation}
\begin{split}
    k_{coll}(T) & = \sigma _{coll}(T) \langle v(T) \rangle \\
    & = \left[
        \pi\left( \dfrac 
            {2C_{6}}
            {k_{B}T}
        \right) ^{1/3}
        \Gamma\left(\dfrac
            {2}
            {3}
        \right)
    \right]
    \left[
        \left( \dfrac 
            {8k_{B}T}
            {\pi\mu}
        \right) ^{1/2}
    \right],
    \label{eqn:kcoll}
\end{split}
\end{equation}

\noindent where \textnormal{$k_{B}$ is the Boltzmann constant, }$\Gamma(x)$ is the gamma function such that $\Gamma(2/3) = 1.353$, $\mu$ is the reduced mass of the collision, and $C_{6}$ is the sum of coefficients describing the magnitude of the attractive forces between collision partners (J cm$^{6}$) \citep{smith1980kinetics,stoecklin1991rate}.  
$C_{6}$ can be described by:
\begin{equation}
    C_{6} = 
    C_{6}^{D-D} +
    C_{6}^{D-iD} +
    C_{6}^{Disp},
    \label{eqn:C6}
\end{equation}

\noindent with $C_{6}^{D-D}$ described by:
\begin{equation}
    C_{6}^{D-D} = 
    \dfrac
        {2}
        {3} 
    \left( \dfrac
        {\mu_{1}^{2}\mu_{2}^{2}}
        {k_{B}T(4\pi\epsilon_{0})^{2}} 
    \right),
    \label{eqn:C6(D-D)}
\end{equation}

\noindent where $\mu_{1}$ and $\mu_{2}$ are the dipole moments of reagents 1 and 2 and $\epsilon_{0}$ is the permittivity of free space \citep{hirschfelder1964molecular}.  
$C_{6}^{D-iD}$ can be described by:
\begin{equation}
    C_{6}^{D-iD} = 
    \dfrac 
        {\mu_{1}^{2}\alpha_{2} + \mu_{2}^{2}\alpha_{1}}
        {4\pi\epsilon_{0}},
    \label{eqn:C6(D-iD)}
\end{equation}

\noindent where $\alpha_{1}$ and $\alpha_{2}$ are the polarizabilities of reagents 1 and 2, and  $C_{6}^{Disp}$ is given by:
\begin{equation}
    C_{6}^{Disp} = 
    \dfrac
        {3}
        {2}
    \alpha_{1}\alpha_{2}\left( \dfrac
        {I_{1}I_{2}}
        {I_{1}+I_{2}}
    \right),
    \label{eqn:C6(Disp)}
\end{equation}

\noindent where $I_{1}$ and $I_{2}$ are the ionization energies of reagents 1 and 2.  
A number of other techniques have been \textnormal{applied} when more accurate descriptions of the intermolecular potential have been needed \citep{phillips1992rate}.  
One technique, rotationally Adiabatic Capture (AC) theory, was developed to calculate the long-range $D-D$, and dipole-quadrupole ($D-Q$) intermolecular potential mostly between  diatomic molecules \citep{stoecklin1991rate,clary1993rate,clary1994rate}.  
In AC theory, long-range intermolecular potentials for individual molecular rotational states are calculated and utilized to determine rate coefficients for each rotational state.  
These single quantum state rate coefficients are then averaged over a Maxwell-Boltzmann distribution in order to determine the final temperature-dependent collision rate coefficient. 
Such collision rate coefficients are found to go to zero at 0 K, increase as $\sim T^{1/6}$ above 0 K \textnormal{when only approximately one state is populated} until reaching a maximum value, and then decrease as $\sim T^{-1/6}$ at higher temperatures \textnormal{when a many states are populated}.  
The maximum value was reached at $\sim 1$ K for simple diatom-diatom collisions, but the maximum was found to shift to $\sim$20 K when electronic effects were considered for OH + HBr, a $^{2}\Pi -$ $ ^{1}\Sigma$ system \citep{clary1993rate}.  
For the reaction CH + NH$_{3}$, AC was able to calculate rate coefficients to within a factor of $\sim$2 of measured values below 100 K \citep{stoecklin1995fast}.  
In a somewhat similar technique, Statistical Adiabatic Capture Model (SACM), adiabatic intermolecular potentials are calculated in order to determine capture rate coefficients \citep{quack1974specific,troe1985statistical}.  
Additionally, the technique of long-range E,J-resolved microcanonical Variational Transition State Theory ($\mu$j-VTST) should yield approximately the same low temperature rate coefficients  as the AC and SACM techniques since the centrifugal barrier becomes the dominant transition state in the determination of the rate coefficient when the reaction has reached the collision limit \citep{georgievskii2005long}.  
$\mu$j-VTST was found to agree with experimental measurements to within a factor of $\sim$2 for the reaction of CN + O$_{2}$ and to within a factor of $\sim$5 for the reaction of CH + NH$_{3}$ near 20 K \citep{georgievskii2005long}.  

In this paper the temperature-dependent rate coefficients for the reaction between CH + CH$_{2}$O were measured between 31 - 133 K and fitted with Equation \ref{eqn:kcoll}, the collision limit derived from classical capture theory.  
The \textnormal{effects of employing} the optimized fit was then determined \textnormal{in} model predictions of the abundances of key species in AGB stellar winds and dark interstellar clouds.

\FloatBarrier
%%%%%%%%%%%%%%%%%%%%%%%%%%%%%%%%%%%%%%
%%EXPERIMENTAL METHOD
%%%%%%%%%%%%%%%%%%%%%%%%%%%%%%%%%%%%%%
\section{Experimental Method} \label{sec:ExpMethod}
\begin{figure}[ht]
    \plotone
        {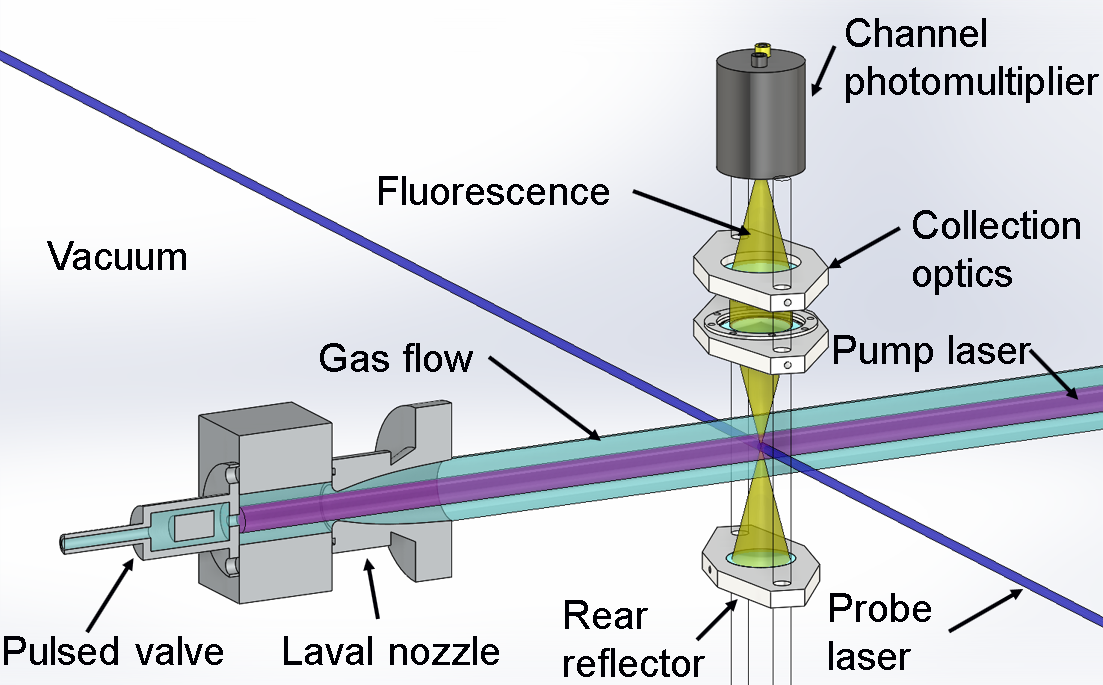}
    \caption{
        Schematic of the pulsed Laval nozzle apparatus and the PLP-LIF technique.  
        Reprinted and adapted with permission from Heard, D. E. 2018, AcChR, 51, 2620. Copyright 2018 American Chemical Society.
        \label{fig:Laval_PLP-LIF_Schematic}
    }
\end{figure}

Low temperature kinetics measurements of the reaction CH + CH$_{2}$O were performed in a pulsed Laval nozzle apparatus, with the Pulsed Laser Photolysis - Laser Induced Fluorescence (PLP-LIF) technique, schematically shown in Figure \ref{fig:Laval_PLP-LIF_Schematic}.  
The reagent and CH precursor gases, formaldehyde and bromoform (CHBr$_{3}$) respectively, were prepared separately before being controllably mixed and directed to the Laval nozzle apparatus.  
Mixtures of formaldehyde and bath gases were prepared in cylinders, utilizing a similar method to those utilized in previous literature \citep{sivakumaran2003reaction,oliveira2016photoelectron}.  
Formaldehyde gas was generated by gently heating the polymerized form of formaldehyde, paraformaldehyde powder (Sigma-Aldrich, 95\%), with a heat gun (Steinel, model HL1810S) to $\sim$70$^{\circ}$C in an evacuated 500 mL glass bottle (Duran) modified to be leak-tight and connected to a vacuum line.  
The formaldehyde gas was passed through a cold trap which was submerged in ethanol (VWR, 99.96\%) chilled to $-10^{\circ}$C with a refrigerated immersion probe (LaPlant, model 100CD) in order to trap any water or other condensable byproducts.  
Empty cylinders attached to the vacuum line were then filled with the purified formaldehyde gas to $\sim$200 Torr ($\sim$26.7 kPa) and then argon (BOC, 99.998\%) or nitrogen (BOC, 99.998\%) gas was added to $\sim$6 atm ($\sim$608 kPa) creating $\sim$4.4\% mixtures of formaldehyde gas.  
The cylinders were then left for $>$12 hours to allow for mixing of the gases.  
In order to generate the precursor bromoform gas, liquid bromoform (Aldrich, 99+\%) was added to a bubbler and the known vapor pressure of bromoform ($\sim$5 Torr ($\sim$0.7 kPa) at 298 K) was entrained in $\sim$2 atm ($\sim$203 kPa) of bath gas \citep{Simnikov1941Increase,linstrom2001nist}.

Individual gases were combined in a mixing manifold with Mass Flow Controllers (MFC) (MKS, type 1179A) such that the final range of experimental gas compositions to be sent to the Laval nozzle apparatus was $\sim$0.1 - 1.0\% CH$_{2}$O, $\sim$0.01\% CHBr$_{3}$, and $\sim$99\% Ar or N$_{2}$ bath gas.  
The absolute concentration of formaldehyde in each final mixture was also determined via UV absorption measurements, since formaldehyde gas slowly reformed solid polymers which sufficiently coated the walls of the MFC  to render the calibration of the flow rate unusable within a day of measurements.  
The gas mixtures were sampled from the tubing between the mixing manifold and the pulsed valves and measured in a custom-built 1 m path length UV absorption cell.  
The absorption light source was a UVB lamp (EXOTERRA, UVB200) with continuous output between $\sim$290 - 350 nm.  
Absorption spectra collected from the UV/Vis spectrometer (Ocean Optics, HR4000CG-UV-NIR) with 0.75 nm resolution were integrated for 2 seconds and 4 spectra traces were averaged in order to generate an averaged spectrum that was utilized to determine the concentration of formaldehyde for each gas mixture.  
The pressure of each gas mixture in the absorption cell was measured by a capacitance manometer (MKS, 0 - 100 PSIA (0 - 689 kPa)) to be $\sim$1.2 atm ($\sim$122 kPa)\textnormal{, approximately equal to the pressure behind the pulsed valves from where the gas mixtures were sampled}.  
Representative UV absorption measurements of CH$_{2}$O are shown in the appendix.  
It was found that the half-life of formaldehyde in a cylinder of initially $\sim$4.4\% formaldehyde in nitrogen was about 4 days.  

After the gas ballast, each final mixture of gas was pulsed at 5 Hz through 2 solenoid valves (Parker, series 9) into the 1 cm$^{3}$ pre-expansion reservoir of the Laval nozzle apparatus.  
The use and characterization of this apparatus was described in detail previously \citep{taylor2008pulsed,shannon2010observation,caravan2014measurements,gomez2014low,shannon2014combined} so only a brief description is given here.  
Each pulse of gas underwent a controlled expansion through a custom-made, axisymmetric, Laval nozzle into a vacuum chamber at 0.3 - 2 Torr (40 - 267 Pa) resulting in a thermalized cold gas flow.  
A range of nozzles with Mach numbers between 5.00 and 2.49 were \textnormal{employed} during the experiments to achieve flow temperatures between 30 - 134 K.  
The density and temperature of the flows were verified with impact pressure measurements as well as by fits of CH LIF rotational temperature measurements to spectra generated in the simulation software package LIFBASE \citep{luque1999lifbase}.  
Each pulse of gas was evacuated from the vacuum chamber by two Roots blower vacuum pump systems in parallel: a Roots blower (Leybold RUVAC 251) backed by a rotary pump (Leybold D65B) and a Roots blower (Edwards EH250) backed by a rotary pump (Edwards ED660).  
The pressure in the vacuum chamber was monitored by a capacitance manometer (Leybold, type CTR90, 0 - 10 Torr (0 - 1.3 kPa)). 

Kinetics experiments in the cold gas flow were carried out with the PLP-LIF technique.  
In order to initiate the reaction of CH with CH$_{2}$O, the CHBr$_{3}$ precursor was photolyzed co-linearly with the nozzle axis with the output of an excimer laser (Lambda Physik, LPX200) at 248 nm, the ``Pump laser" in Figure \ref{fig:Laval_PLP-LIF_Schematic}, generating a uniform density of CH radicals.  
\begin{subequations} %CH + CH2O Eqns
    \label{eqn:CHBR3_Photolysis}
\begin{align}
        \ce{&CHBr3 + 3h$\nu_{248 nm}$ -> CH + \textnormal{Br + Br$_{2}$}},
        \label{eqn:CHBR3_Photolysis_To_Ground}\\
        \ce{&CHBr3 + 3h$\nu_{248 nm}$ -> CH$^{*}$ + \textnormal{Br + Br$_{2}$}},
        \label{eqn:CHBR3_Photolysis_Excited}
\end{align}
\end{subequations}

\begin{equation}
    \textnormal{\ce{CH$^{*}$ ->[$k_{rel}^{emission}$] CH + h$\nu$},}
    \label{eqn:Excited_CH_Emission}%\\
\end{equation}

\begin{equation}
    \textnormal{\ce{CH$^{*}$ + Q ->[$k_{rel}^{quench}$] CH + Q$^{*}$},}
    \label{eqn:Excited_CH_Relaxation}
\end{equation}

\begin{equation}
    k_{rel}^{\prime} = k_{rel}^{emission} + k_{rel}^{quench}[Q],
    \label{eqn:k_rel_prime}
\end{equation}

\noindent where CH$^{*}$ is rotationally, vibrationally, and/or electronically excited CH and $k_{rel}^{\prime}$ is a simplified first-order rate coefficient approximating many pathways of electronic, vibrational, and rotational relaxation of CH$^{*}$ to CH, and Q is any species which collisionally relaxes CH$^{*}$\textnormal{ \citep{lindner1998multi,zou2004photodissociation}}.  
The transient relative concentration of CH radicals was then monitored via the Q$_{2}$(1) rotational line of the $B^{2}\Sigma - X^{2}\Pi$ (1,0) vibronic transition at 363.569 nm by probing with a pulsed Nd:YAG (Litron, LYP 664-10) pumped dye laser (Sirah, Cobra Stretch), the ``Probe laser" in Figure \ref{fig:Laval_PLP-LIF_Schematic}.  
The probe laser beam was passed transversely through the gas flow, perpendicularly crossing the pump laser at the furthest distance from the exit of the nozzle before the flow broke up due to turbulence (typically $\sim$10 - 25 cm depending on the nozzle).  
The resulting fluorescence was focused with a series of lenses through two optical filters: a bandpass Filter at 400 nm with a Full Width at Half Max (FWHM) of 40 nm (Thorlabs, FB400-40) and a clear acrylic $\sim$400 nm long-pass filter (Perspex)\textnormal{ removing CH(B-X)(0,0) emission at $\sim390$ nm and selecting the CH(B-X)(1,1) at $\sim404$ nm in order to minimize CH* emission from CHBr$_{3}$ photolysis at pump-probe time delays at $\lesssim1\mu s$}.  
The filtered fluorescence was then collected by a temporally gated Channel PhotoMultiplier (CPM) (PerkinElmer, C1952P) with spectral response over 165-750 nm.  
The signal from the CPM was then digitized and integrated on an oscilloscope (LeCroy, Waverunner LT264) and the integrated signal was sent to a custom LabVIEW program. 
This LabVIEW program also controlled a digital delay generator (BNC, Model 555) which controlled the timing of the experiment.

\FloatBarrier
%%%%%%%%%%%%%%%%%%%%%%%%%%%%%%%%%%%%%%
%%EXPERIMENTAL RESULTS AND DISCUSSION
%%%%%%%%%%%%%%%%%%%%%%%%%%%%%%%%%%%%%%
\section{Experimental Results And Discussion} \label{sec:ExpResults}

\begin{figure}[ht] %CH_LIF_vs_t_traces
    \plotone
        {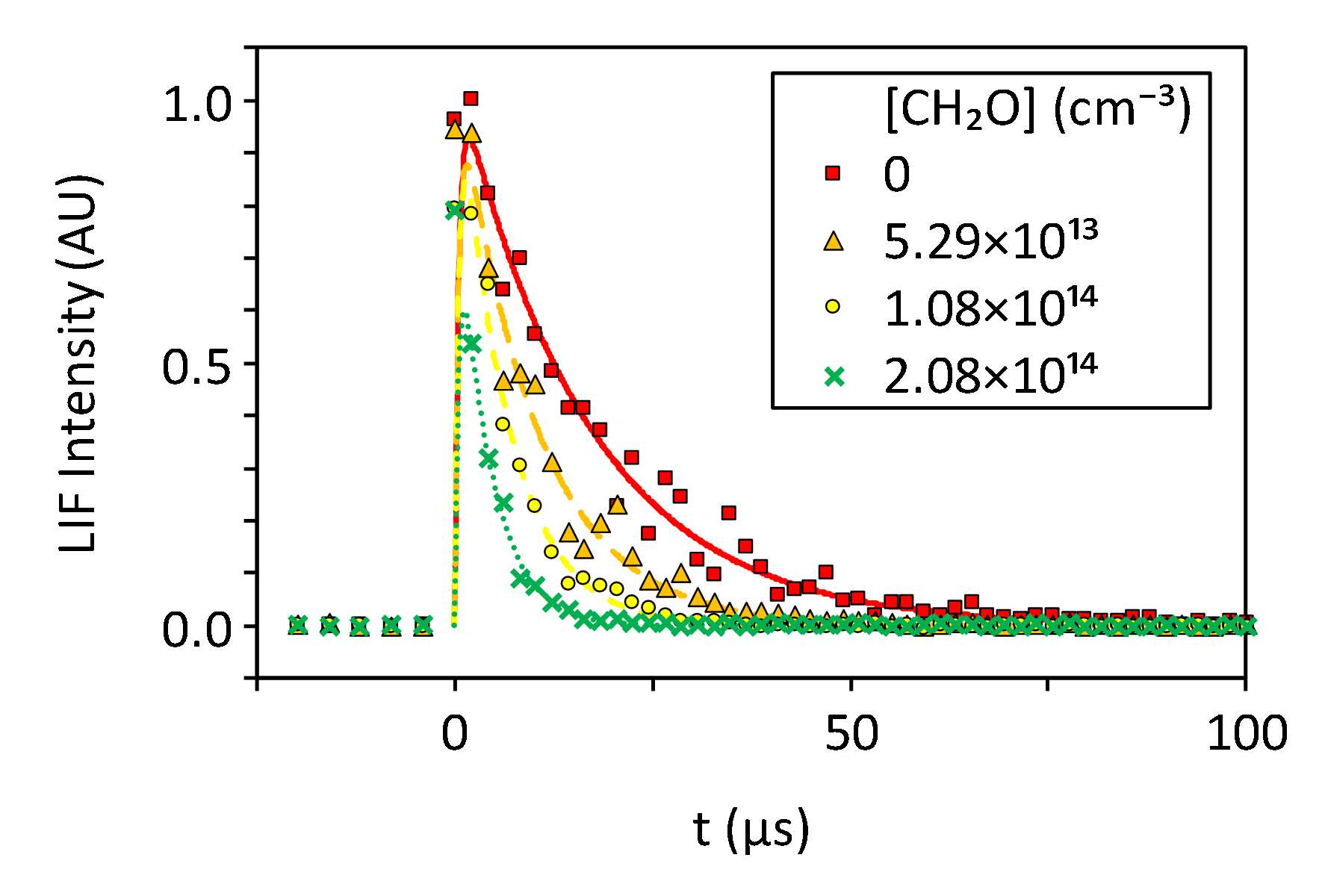}
    \caption{
        Representative transient CH integrated LIF traces utilized to determine the pseudo-first-order rate coefficients for loss of CH at 133 K and [N$_{2}$] $= 1.07\times 10^{17}$ cm$^{-3}$, together with biexponential fits to the data for various concentrations of CH$_{2}$O.  
        \label{fig:CH_LIF_vs_t_traces}
    }
\end{figure}

In order to determine the temperature-dependent rate coefficients for the loss of CH due to reaction with CH$_{2}$O, the pseudo-first-order rate coefficients, $k_{obs}$, were first measured at multiple concentrations of CH$_{2}$O for a given temperature.  
The relative temporal evolution of CH was monitored by integrating CH LIF while randomly varying the pump-probe time delay for each gas pulse up to the longest time delay in which nascent CH, generated in the flow at the exit of the nozzle, took to reach the probed region of the flow ($\sim$100 - 300 $\mu$s depending on the nozzle).  
This process was repeated so that the CH signal at each time delay was averaged from at least 4 laser shots.  
Representative traces of integrated CH LIF versus time for several CH$_{2}$O concentrations are shown in Figure \ref{fig:CH_LIF_vs_t_traces}.  

In each CH trace, following the instantaneous production via Reaction \ref{eqn:CHBR3_Photolysis_To_Ground}, there was a fast rise ($\lesssim$5 $\mu$s) due to  relaxation of excited CH*, formed by Reaction \ref{eqn:CHBR3_Photolysis_Excited}, into the rovibrational level of the X$^{2}\Pi$ ground state probed via LIF, followed by an exponential decay due to both the diffusion of CH out of the probe laser beam volume:
\begin{equation}
    \ce{CH ->[$k_{diff}$] (diffusive loss)},
    \label{eqn:CH_Diffusion}
\end{equation}
and reaction of CH primarily with CH$_{2}$O, Reaction \ref{eqn:Rxn_CH_CH2O-Prod}, and also CH with other species:
\begin{equation}
    \ce{CH + X$_{i}$ ->[$k^{other}_{i}$] products},
    \label{eqn:CH_Other}
\end{equation}
where X$_{i}$ is each non-reagent species $i$: namely N$_{2}$ (when N$_{2}$ was present as the bath gas), and also the precursor CHBr$_{3}$\textnormal{, one and two-photon photolysis products of CHBr$_{3}$, and as stated by the manufacturer, the CHBr$_{3}$ stabilizer 2-methyl-2-butene, which was present at 60-120 ppm in the CHBr$_{3}$ liquid}.  
The observed pseudo-first-order rate coefficient for the loss of CH, $k_{obs}$, is then:
\begin{equation}
\begin{split}
    k_{obs}&=k_{1}[\textrm{CH}_{2}\textrm{O}] + k_{diff} + \sum ^{N} _{i=1} \left( [\textrm{X}_{i}]k^{other}_{i}\right)\\
    &=k_{1}[\textrm{CH}_{2}\textrm{O}] + k_{int},
    \label{eqn:kobs}
\end{split}
\end{equation}

\noindent where $N$ is the total number of non-reagent species in the cold gas flow.  
For each CH trace, the average background integrated LIF signal was determined by averaging the integrated LIF signal at negative pump-probe time delays.  
Each CH trace was then corrected by subtracting the average background signal, and then fitted utilizing a biexponential function, given by Equation \ref{eqn:Biexpfit}, where the observed pseudo-first-order rate coefficients of the exponential rise ($k_{rel}^{\prime}$) and exponential decay ($k_{obs}$) were the fitted parameters and allowed to vary from trace to trace:  
\begin{equation}%BiexponentialFit
    I(t) 
    = f_{1}e^{-k_{obs}t}-f_{2}e^{-k_{rel}^{\prime}t}
    \propto [CH]_{t},
\end{equation}

\noindent where

\begin{equation}%BiexponentialFit
    [CH]_{t} = \left[
        [CH]_{t=0} + \dfrac
            {k_{rel}^{\prime}}
            {k_{rel}^{\prime}-k_{obs}}
        [CH^{*}]_{t=0}
    \right]
    e^{-k_{obs}t}
    - \left[ 
        \dfrac
            {k_{rel}^{\prime}}
            {k_{rel}^{\prime}-k_{obs}}
        [CH^{*}]_{t=0}
    \right]
    e^{-k_{rel}^{\prime}t},
    \label{eqn:Biexpfit}
\end{equation}

\begin{figure*}[ht] %Intercept_Subtracted_Second_Order_Plots
    \plottwo
        {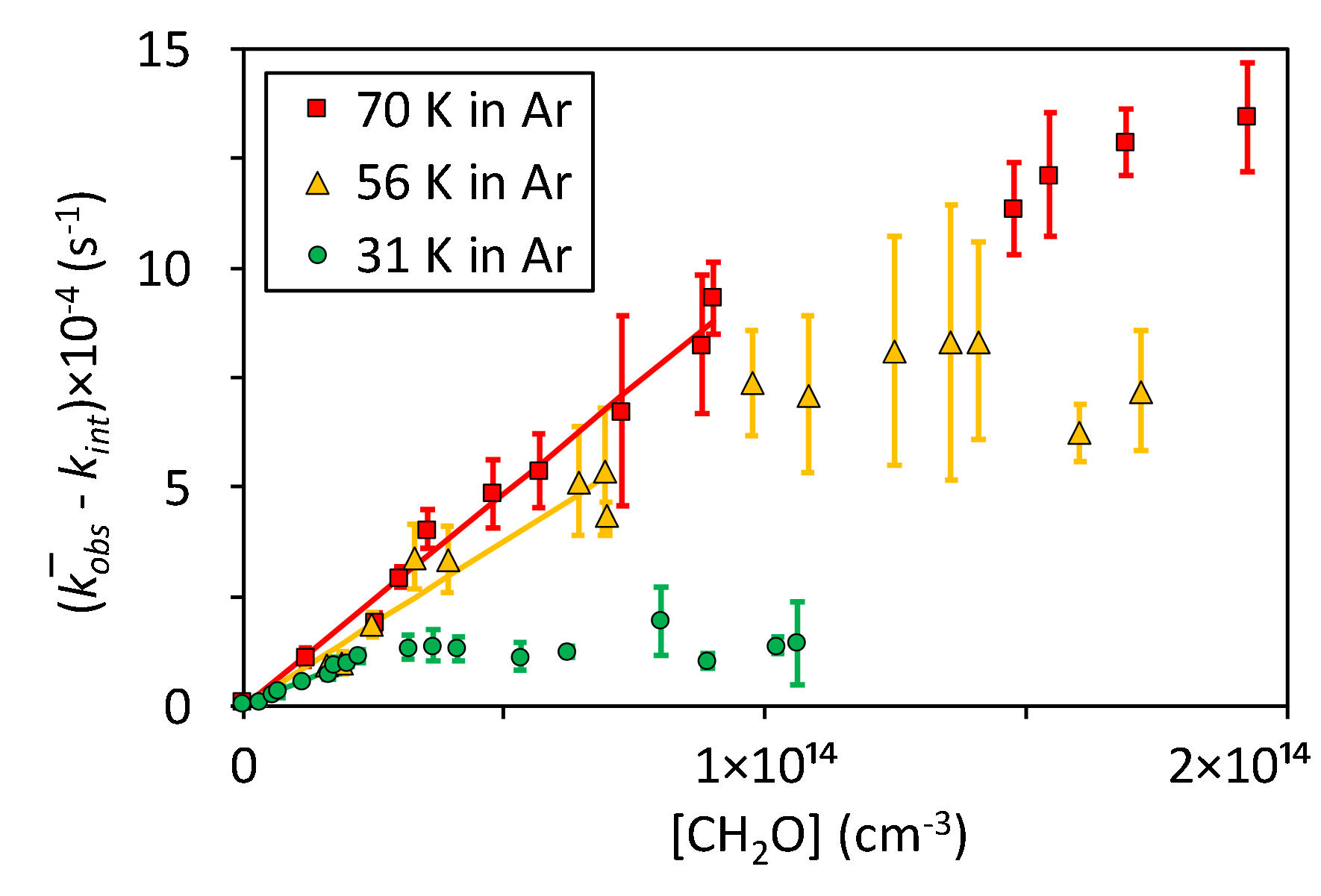}
        {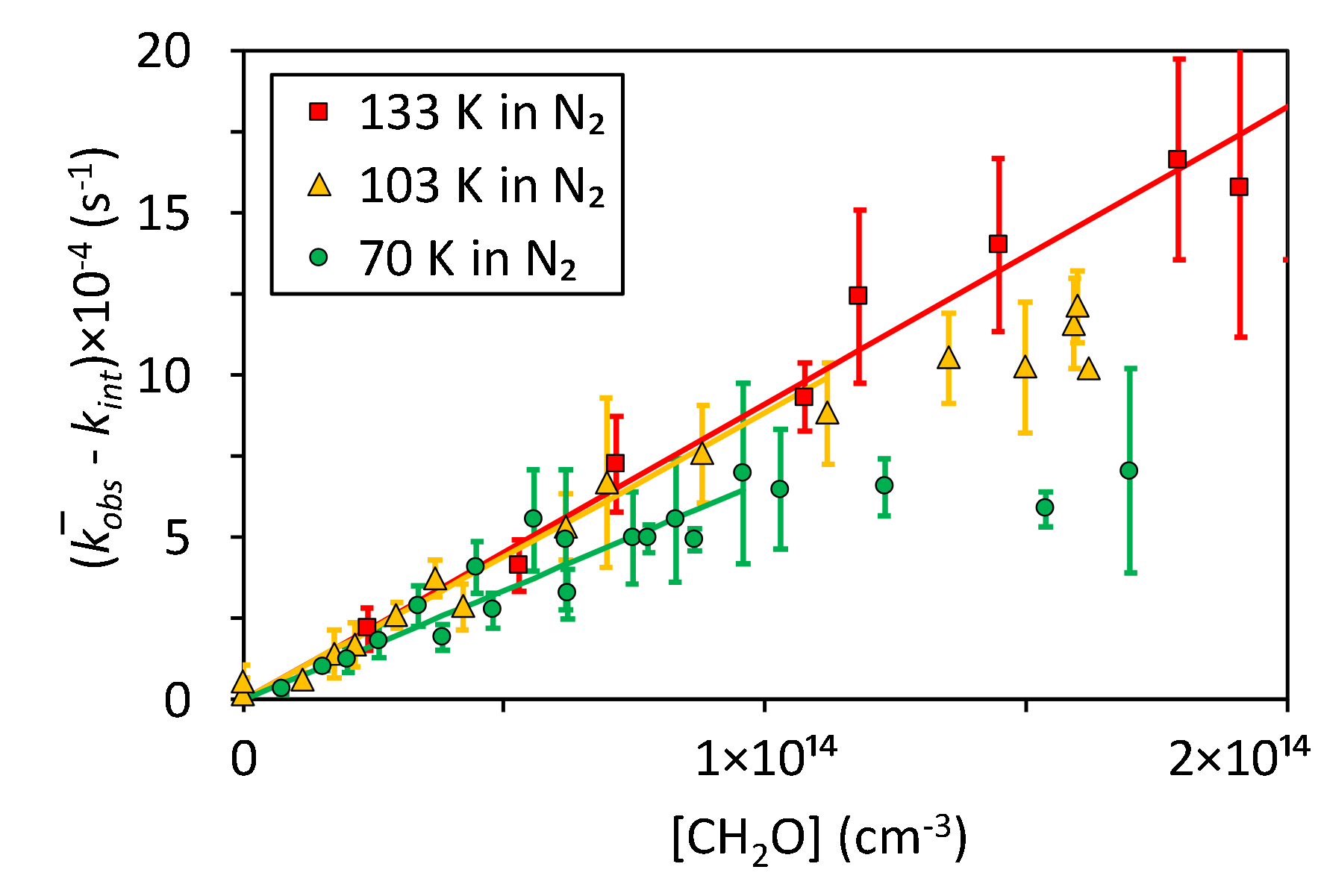}
    \caption{
        Intercept-subtracted average rate of loss of CH (Equations \ref{eqn:kobs} and \ref{eqn:k_obs_bar}) versus the concentration of formaldehyde at various temperatures along with linear fits utilizing Equation \ref{eqn:kobs} at each temperature.  
        Error bars represent one standard deviation of fits of $k_{obs}$ from at least 5 experimental CH temporal traces.  
        The corresponding plots without subtraction of the \textnormal{$k_{int}$ values} are shown in Figure \ref{fig:Nonsub_Second_Order_Plots} of Appendix \ref{sec:Second_Order_Plots}.  
        \label{fig:Int_Sub_Second_Order_Plots}
    }
\end{figure*}

\noindent $I(t)$ is the time dependent LIF signal, [CH]$_{t=0}$ and [CH$^{*}$]$_{t=0}$ are the initial concentrations of CH and CH$^{*}$, and $f_{1}$ and $f_{2}$ are fitted constants since the initial concentrations of [CH]$_{t=0}$ and [CH$^{*}$]$_{t=0}$ as well as the values of $k_{rel}^{\prime}$ are not all known under the experimental conditions in this work.  
The values of $k_{obs}$ obtained from biexponential fits were also compared to values obtained by fitting single exponential decay curves starting after $\sim$20 $\mu$s in the experimental traces, and these $k_{obs}$ values were equivalent to those of the biexponential fits to within statistical significance.  
Each averaged CH trace was re-collected at least 5 times, and the fits of $k_{obs}$ for these traces were averaged to obtain a $\bar{k}_{obs}$ value for each [CH$_{2}$O].
\begin{equation}%BiexponentialFit
    \bar{k}_{obs} = \dfrac
        {1}
        {N}
    \sum ^{N}_{i=1}k_{obs,i},
    \label{eqn:k_obs_bar}
\end{equation}

\noindent where $N$ is the number of fit traces and $k_{obs,i}$ is each fit $i$.  The average values of fits of the rate of loss of CH, $\bar{k}_{obs}$, minus the intercept, $k_{int}$ in Equation \ref{eqn:kobs}, versus [CH$_{2}$O] are shown in Figure \ref{fig:Int_Sub_Second_Order_Plots}.  
The fitted value of \textnormal{$k_{int}$} for each second order plot was between 8,000 $\lesssim$ $k_{int}$ (s$^{-1}$) $\lesssim$ 60,000, shown in Figure \ref{fig:Nonsub_Second_Order_Plots} of Appendix \ref{sec:Second_Order_Plots}.  \textnormal{Since all values of $k_{int}$ in Ar were $\lesssim8,000$ s$^{-1}$, the reaction CH + N$_{2}$ was estimated to account for up to $\sim$52,000 s$^{-1}$ of $k_{int}$.}  

\begin{figure}[ht] %k1(T)_vs_T_Measurements
    \plotone
        {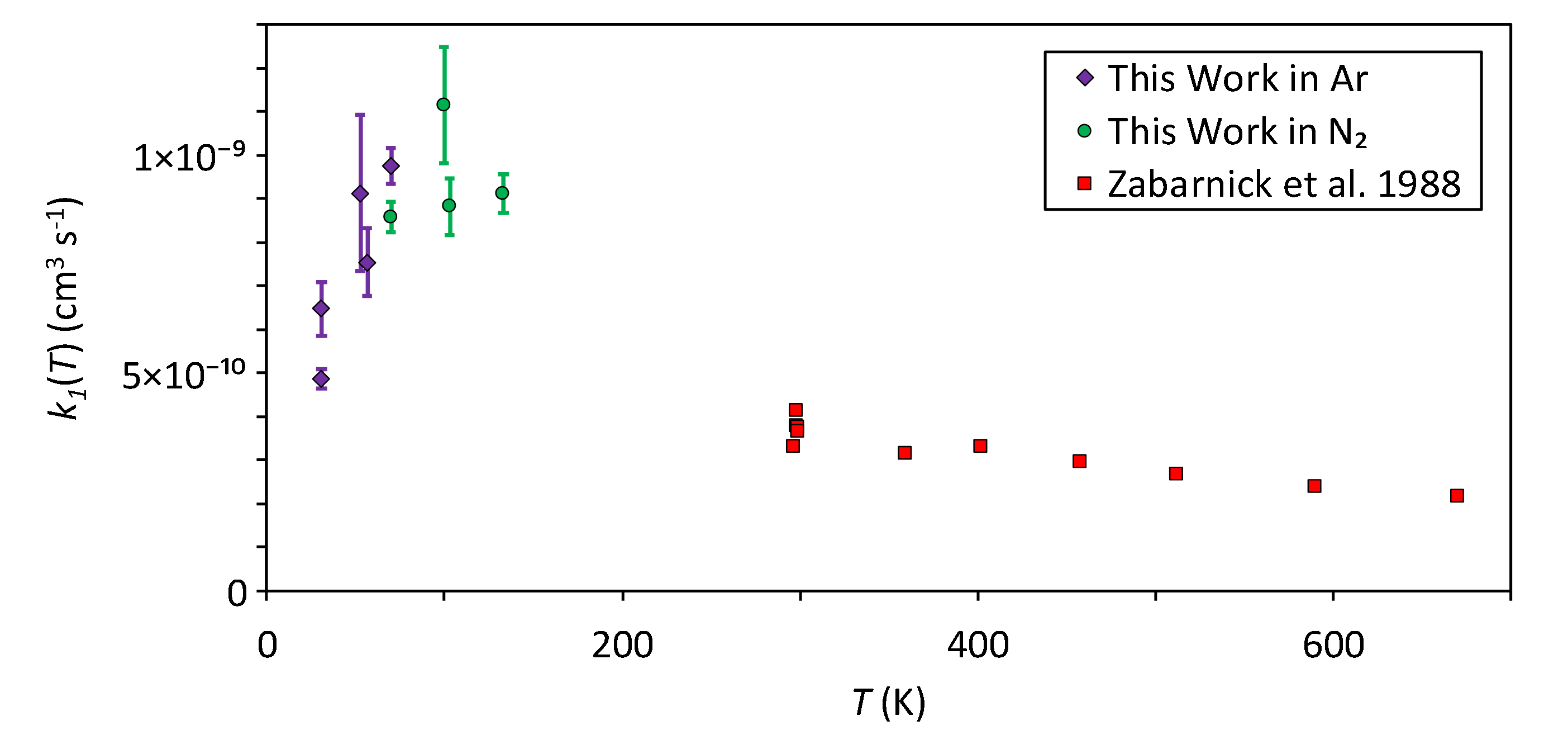}
    \caption{
        Measured values of $k_{1}(T)$ versus temperature.  
        The error bars of each value of $k_{1}(T)$ from this work represent the error in the fit\textnormal{ted} slope of each second-order plot and do not include systematic errors.
        \label{fig:k1(T)_vs_T_Measurements}
    }
\end{figure}

For a given temperature, $\bar{k}_{obs}$ values increased linearly with [CH$_{2}$O] until the formation of formaldehyde dimers, (CH$_{2}$O)$_{2}$, and higher order oligomers, (CH$_{2}$O)$_{n>2}$, began to occur, which resulted in the curving over of the second order plots of $\bar{k}_{obs}$ vs [CH$_{2}$O] at the largest [CH$_{2}$O] values (i.e. the slowing or cessation of the increase in $\bar{k}_{obs}$ with [CH$_{2}$O]).  
The negative curvature of $\bar{k}_{obs}$ at higher [CH$_{2}$O] implies that formaldehyde dimers do not react fast enough with CH to counterbalance the loss of CH$_{2}$O monomers, and hence $k_{obs}$, due to dimerization.  
Therefore, the linear fits of $\bar{k}_{obs}$ versus [CH$_{2}$O] only included $\bar{k}_{obs}$ values from experiments where [CH$_{2}$O] was low enough such that no significant dimerization had occurred, where there was no significant curvature of $\bar{k}_{obs}$ versus [CH$_{2}$O].  
The slopes of the linear fits represent the rate coefficient, $k_{1}(T)$, for Reaction \ref{eqn:Rxn_CH_CH2O-Prod} at a given temperature.  
Values of $k_{1}(T)$ with the corresponding experimental conditions are shown in Table \ref{tab:Measured_Rate_Coefs}.  
\textnormal{Measurements of $k_{1}(T)$ were repeated at 31 K and 70 K in order to verify experimental reproducibility. }
Values of $k_{1}(T)$ were found to have a positive temperature dependence over the range 31 $<$ T $<$ 133 K, and were also found to be independent of pressure at 70 K over a factor of 4.3 change in bath gas density (either N$_{2}$ or Ar) from $N_{total} = 2.58\times10^{16} - 11.18\times10^{16}$ cm$^{-3}$.  
Fit values of $k_{1}(T)$ versus $T$ are shown in Figure \ref{fig:k1(T)_vs_T_Measurements} including measurements from \cite{zabarnick1988temperature} over the temperature range 300 - 670 K. 
Experiments from \cite{zabarnick1988temperature} measured the partial pressure of CH$_{2}$O manometrically before mixing gases in order to determine each [CH$_{2}$O], but did not use UV absorption to determine each [CH$_{2}$O].  

    \begin{deluxetable}{cccc}
    \tablecaption{
        Rate coefficients and experimental conditions for CH + CH$_{2}$O.    
        \label{tab:Measured_Rate_Coefs}
    }
    \tablehead{
        %%%%%%%%%%%%%%%%%%%%%%%%%%%%%%%%%%%%%%%%%%%%%%%%%%
        \colhead{T} & \colhead{Bath Gas} & \colhead{N$_{total}$} &    \colhead{$k_{1}(T)$} \\
        %%%%%%%%%%%%%%%%%%%%%%%%%%%%%%%%%%%%%%%%%%%%%%%%%%
        \colhead{($K$)} & \colhead{} & \colhead{$(10^{16}$ cm$^{-3})$} & \colhead{$(10^{-10}$ cm$^{3}s^{-1})$} 
        %%%%%%%%%%%%%%%%%%%%%%%%%%%%%%%%%%%%%%%%%%%%%%%%%%
    }
    %\colnumbers
    \startdata
        %%%%%%%%%%%%%%%%%%%%%%%%%%%%%%%%%%%%%%%%%%%%%%%%%%
        31 $\pm$ 2 & Ar & 3.24 $\pm$ 0.24 & 4.86 $\pm$ 0.22 \\
        %%%%%%%%%%%%%%%%%%%%%%%%%%%%%%%%%%%%%%%%%%%%%%%%%%
        31 $\pm$ 2 & Ar & 3.24 $\pm$ 0.24 & 6.48 $\pm$ 0.62 \\
        %%%%%%%%%%%%%%%%%%%%%%%%%%%%%%%%%%%%%%%%%%%%%%%%%%
        53 $\pm$ 4 & Ar & 7.04 $\pm$ 0.74 & 9.13 $\pm$ 1.80 \\
        %%%%%%%%%%%%%%%%%%%%%%%%%%%%%%%%%%%%%%%%%%%%%%%%%%
        57 $\pm$ 8 & Ar & 8.14 $\pm$ 1.67 & 7.55 $\pm$ 0.79 \\
        %%%%%%%%%%%%%%%%%%%%%%%%%%%%%%%%%%%%%%%%%%%%%%%%%%
        70 $\pm$ 11 & Ar & 11.18 $\pm$ 2.54 & 9.75 $\pm$ 0.42 \\
        %%%%%%%%%%%%%%%%%%%%%%%%%%%%%%%%%%%%%%%%%%%%%%%%%%
        70 $\pm$ 2 & N$_{2}$ & 2.58 $\pm$ 0.14 & 8.59 $\pm$ 0.35 \\
        %%%%%%%%%%%%%%%%%%%%%%%%%%%%%%%%%%%%%%%%%%%%%%%%%%
        100 $\pm$ 6 & N$_{2}$ & 6.03 $\pm$ 0.84 & 11.15 $\pm$ 1.34 \\
        %%%%%%%%%%%%%%%%%%%%%%%%%%%%%%%%%%%%%%%%%%%%%%%%%%
        103 $\pm$ 10 & N$_{2}$ & 6.80 $\pm$ 1.57 & 8.82 $\pm$ 0.66 \\
        %%%%%%%%%%%%%%%%%%%%%%%%%%%%%%%%%%%%%%%%%%%%%%%%%%
        133 $\pm$ 13 & N$_{2}$ & 10.70 $\pm$ 3.50 & 9.13 $\pm$ 0.45 \\
        %%%%%%%%%%%%%%%%%%%%%%%%%%%%%%%%%%%%%%%%%%%%%%%%%%
    \enddata
    \tablecomments{
        The error of each $k_{1}(T)$ value represents the error in the fitted value of the slope of $\bar{k}_{obs}$ versus [CH$_{2}$O] and does not include systematic errors.  
        The errors in each value of $T$ and $N_{total}$ were calculated by first taking pitot pressure measurements in the cold flow along the axis of the nozzle, converting these values to temperature and density using thermodynamic relations, and then taking the standard deviation of these values.  
    }
    \end{deluxetable}

\begin{figure}[hb] %k1(T)_and_kcoll(T)_vs_T
    \plotone
        {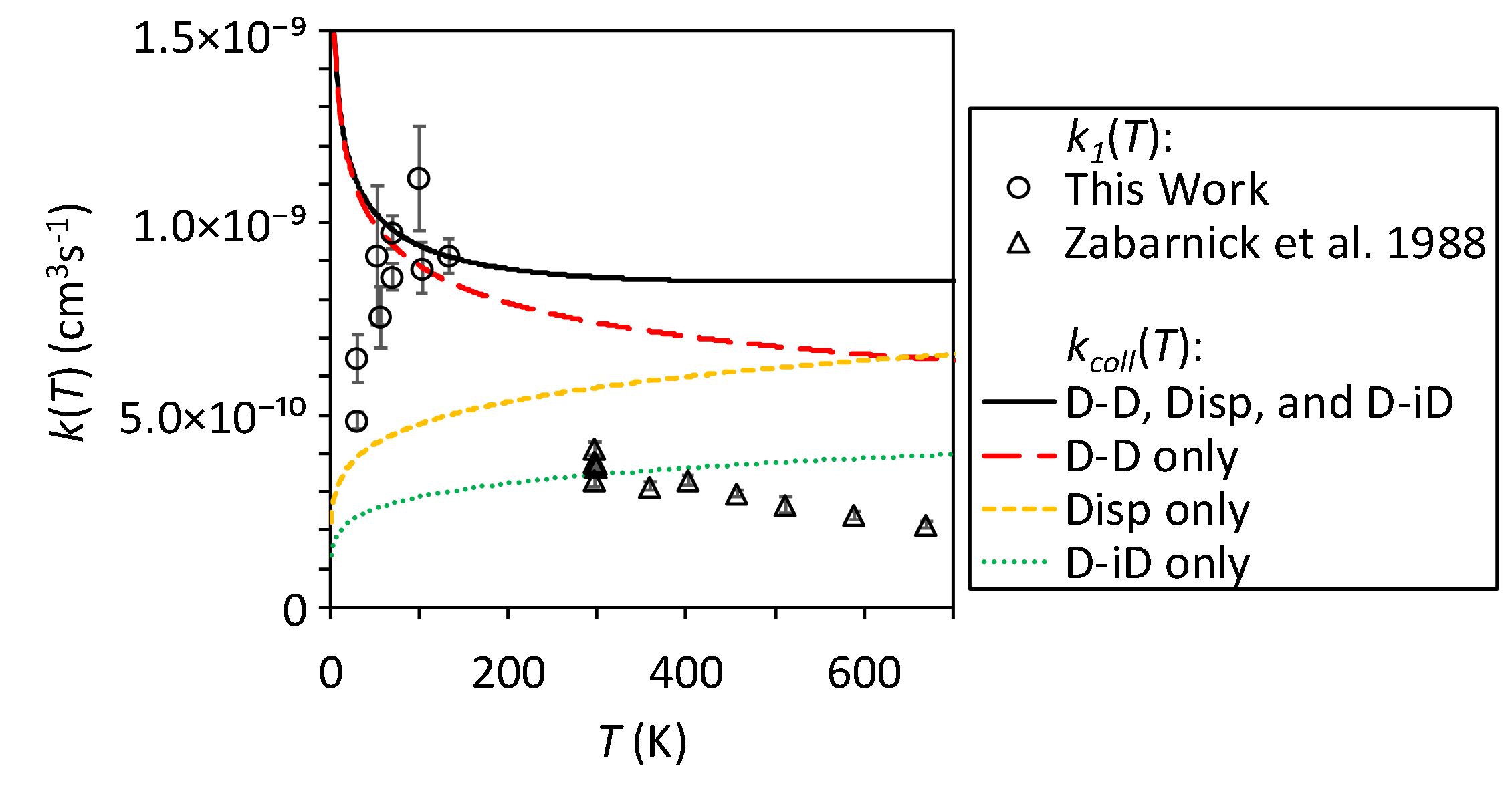}
    \caption{
        Measured values of $k_{1}(T)$ versus temperature compared with the calculated $k_{coll}(T)$ curve, Equation \ref{eqn:kcoll}, and various relative contributions toward $k_{coll}(T)$.  
        For  ``$D-D$ only,"  ``$Disp$ only," and  ``$D-iD$ only" curves, $k_{coll}(T)$ was calculated with $C_{6}$ = $C_{6}^{D-D}$, $C_{6}$ = $C_{6}^{Disp}$, and $C_{6}$ = $C_{6}^{D-iD}$ respectively, Equations \ref{eqn:C6(D-D)}-\ref{eqn:C6(Disp)}, in order to show the relative contribution of each intermolecular force to the total $k_{coll}(T)$.  
        The error bars of each $k_{1}(T)$ value from this work represent the error in the fit value of the slope of each second-order plot and do not include systematic errors.  
        \label{fig:k1(T)_and_kcoll(T)_vs_T}
    }
\end{figure}

\cite{zabarnick1988temperature} observed a mild negative temperature dependence of $k_{1}(T)$ versus $T$ between 298 - 670 K.  
Since a positive temperature dependence of $k_{1}(T)$ between 31 - $\sim$100 K was observed in our experiments, the change from the positive temperature dependence to the negative temperature dependence of $k_{1}(T)$ must occur between $\sim$100 - 298 K, suggesting that the reaction mechanism changes in this range.  
If $k_{coll}(T)$ is calculated using classical capture theory, Equation \ref{eqn:kcoll}, with the constants given in Table \ref{tab:kcoll_parameters}, and plotted versus temperature, as in Figure \ref{fig:k1(T)_and_kcoll(T)_vs_T}, then the values of $k_{coll}(T)$ agree fairly well with the experimental values of $k_{1}(T)$ near 100 K, suggesting that the reaction has reached the collision limit.  

According to the CCT model, shown as a solid black curve in Figure \ref{fig:k1(T)_and_kcoll(T)_vs_T}, values of $k_{1}(T)$ cannot be larger than the limiting values in which every collision between CH and CH$_{2}$O results in a reaction. 
Therefore, the negative temperature dependence of $k_{1}(T)$ observed between 300 - 670 K should not be extrapolated below $\sim$150 K where the collision limit is reached.  
\textnormal{Furthermore, since the linear addition of the strengths of $D-D$, $Disp$, and $D-iD$ forces is somewhat different than the calculation of a long range potential with high level \textit{ab initio} methods, and since there is always some error in experiments, it is not surprising that there are measured $k_{1}(T)$ values that are slightly greater than the first order estimate of the collision limit, $k_{coll}(T)$.  }
At temperatures below $\sim$100 K, calculated values of $k_{coll}(T)$ diverge from measured $k_{1}(T)$ values, likely due to the limitations of classical capture theory, Equation \ref{eqn:kcoll}, in not adequately describing $D-D$ and $D-Q$ interactions at these temperatures.  
When studying similar, diatom-diatom systems with the more rigorous AC approach, \citet{clary1993rate}  observed a temperature dependence of $\sim T^{1/6}$ near 0 K and $\sim T^{-1/6}$ at higher temperatures with a maximum at a temperature (1 $\lesssim$ T (K) $\lesssim$ 40) which varied for reactions with differing ground electronic states.  
While AC theory calculations have been shown to predict similar temperature dependencies as those measured for $k_{1}(T)$ in this work, AC theory has not been able to match experimental rate coefficient values to better than a factor of ~2.  
\textnormal{Additionally, it is possible that some other mechanism is causing $k_{1}(T)$ to have values less than the collision limit below $\sim$100 K.  
However, the mechanisms governing the negative temperature dependence of $k_{1}(T)$ above $\sim$100 K would have led to a further increase in $k_{1}(T)$ at lower temperatures if not for the collision limit.  
A new mechanism would have to explain why $k_{1}(T)$ has a positive temperature dependence below $\sim$100 K when there are strong mechanisms that would cause a negative temperature dependence.  
It is instead more likely that $k_{1}(T)$ is still governed by the collision limit below $\sim$100 K, and that the collision limit has a positive temperature dependence.  }

    \begin{deluxetable*}{cccccc}%kcoll_parameters
        \tablecaption{
            Parameters used to calculate $k_{coll}(T)$ between CH and CH$_{2}$O.
        \label{tab:kcoll_parameters}
    }
    \tablehead{
        %%%%%%%%%%%%%%%%%%%%%%%%%%%%%%%%%%%%%%%%%%%%%%%%%%
        \colhead{Molecule$\quad$} & \multicolumn{2}{c}{Dipole Moment} & \colhead{$\quad$$\quad$Polarizability$\quad$$\quad$} & \multicolumn{2}{c}{Ionization Energy} \\
        %%%%%%%%%%%%%%%%%%%%%%%%%%%%%%%%%%%%%%%%%%%%%%%%%%
        \colhead{} & \multicolumn{2}{c}{$\mu_{n}$} & \colhead{$\alpha_{n}$} & \multicolumn{2}{c}{$I_{n}$} \\
        %%%%%%%%%%%%%%%%%%%%%%%%%%%%%%%%%%%%%%%%%%%%%%%%%%
        \colhead{} & \colhead{$\quad$(Debye)} & \colhead{(C cm)$\quad$} & \colhead{(cm$^{3}$)} & \colhead{(eV)} & \colhead{(J)}
        %%%%%%%%%%%%%%%%%%%%%%%%%%%%%%%%%%%%%%%%%%%%%%%%%%
    }
    %\colnumbers
    \startdata
        %%%%%%%%%%%%%%%%%%%%%%%%%%%%%%%%%%%%%%%%%%%%%%%%%%
        CH & 
        %0.574 $\pm$ 0.023 [1] & %Atomic Units
        1.46 &
        4.87$\times 10^{-28}$ [1] &
        %16.22 [3]& %Atomic Units
        2.40$\times 10^{-24}$ [3]& 
        %10.640 $\pm$ 0.010 [5] \\
        10.640 &
        1.70$\times 10^{-18}$ [5] \\
        %%%%%%%%%%%%%%%%%%%%%%%%%%%%%%%%%%%%%%%%%%%%%%%%%%
        CH$_{2}$O & 
        %0.917 [2] & %Atomic Units
        2.33 &
        7.77$\times 10^{-28}$ [2] &
        %18.69 [4] & %Atomic Units
        2.77$\times 10^{-24}$ [4]& 
        %10.8887 $\pm$ 0.0030 [6] \\
        10.8887 &
        1.74$\times 10^{-18}$ [6] \\
        %%%%%%%%%%%%%%%%%%%%%%%%%%%%%%%%%%%%%%%%%%%%%%%%%%
    %\tablenotemark{a}
    \enddata
    \tablerefs{
        [1]\cite{phelps1966experimental}; 
        [2]\cite{nelson1967selected};
        [3]\cite{manohar2007dipole}; 
        [4]\cite{olney1997absolute}; 
        [5]\cite{herzberg1969new}; 
        [6]\cite{niu1993high}
    }
    \end{deluxetable*}

\begin{figure}[htb!] %k1(T)_and_All_Fits_vs_T
    \plotone
        {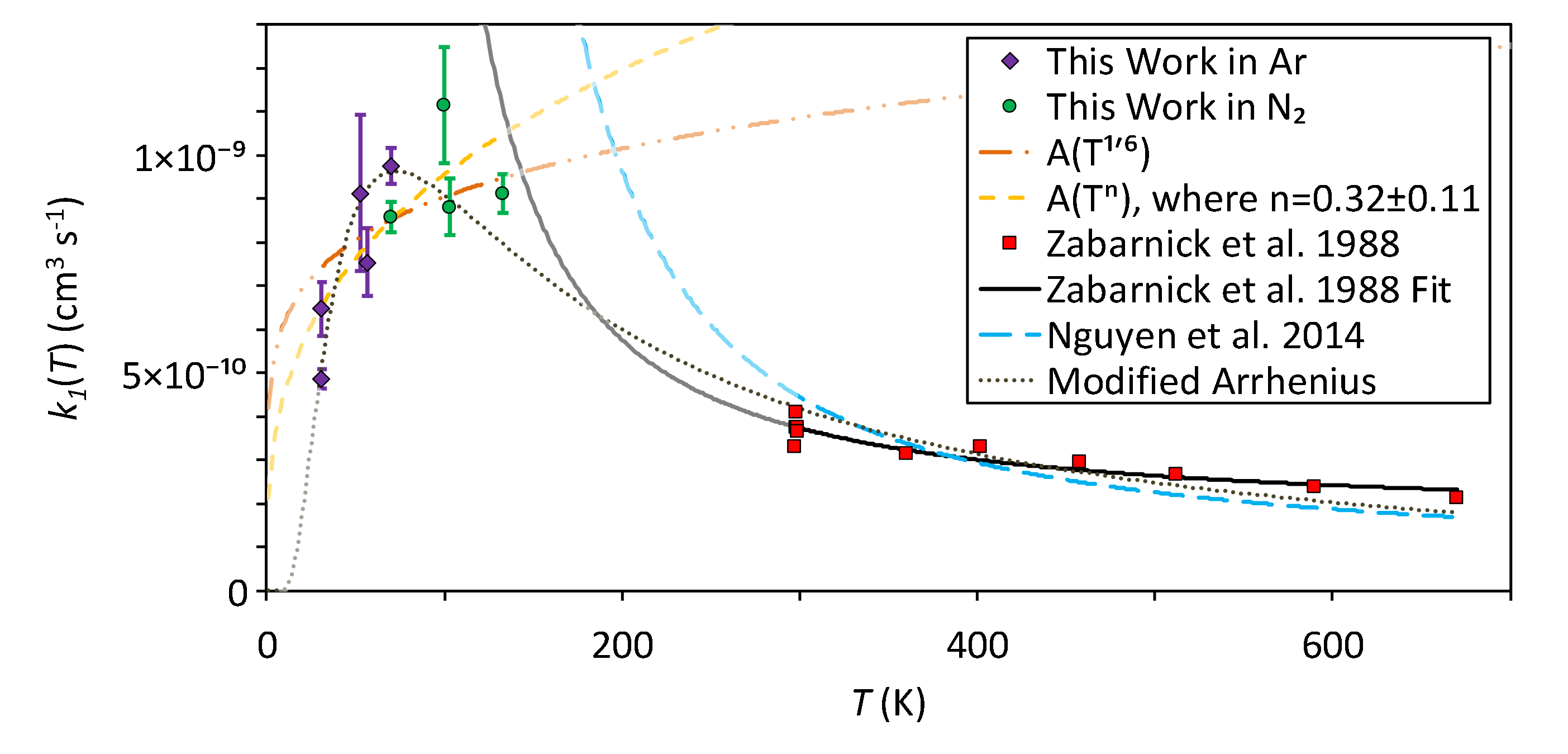}
    \caption{
        Measured values of $k_{1}(T)$ versus temperature together with fits using several approaches.  
        Extrapolations of fit curves beyond measured values of $k_{1}(T)$ are shown with a lightened color of each respective fit curve.  
        The error bars of each $k_{1}(T)$ value from this work represent the error in the fit value of the slope of each second-order plot and do not include systematic errors.  
        The red square data points and the black fit curve were taken from \cite{zabarnick1988temperature} and the blue long dashed curve was taken from \cite{nguyen2014theoretical}.  
        \label{fig:k1(T)_and_All_Fits_vs_T}
    }
\end{figure}

If the fit of the measured negative temperature dependence of $k_{1}(T)$ between 300 - 670 K is extrapolated to 133 K, and then a temperature dependence of $T^{1/6}$ is \textnormal{applied} below 133 K, the experimental data are reasonably represented as shown in Figure \ref{fig:k1(T)_and_All_Fits_vs_T} by the dotted brown curve.  
Furthermore, the total range of measured values of $k_{1}(T)$ between 31 - 670 K could be fit to within 28\% to the modified Arrhenius equation, Equation \ref{eqn:ModArrh}.  
Note that this fit should not be used for prediction of the rate coefficient outside of the range of temperatures of 31 - 670 K.  
If instead of following an extrapolation of the modified Arrhenius curve, the rate coefficients were to follow a $T^{1/6}$ temperature dependence below 30 K, as predicted by adiabatic capture theory, the value given by the modified Arrhenius equation, Equation \ref{eqn:ModArrh}, would be incorrect by a factor of $\sim$140 at 10 K.  
Additionally, an $A\times T^{n}$ fit was performed of the data between 31 - 133 K which indicated that the measured positive temperature dependence ($n = 0.32 \pm 0.11$) in this temperature range may be more appropriate than a $T^{1/6}$ dependence.  
The $A\times T^{n}$ fits are consistent with the form of the temperature dependence of the collision rate coefficients at the limit of $T\rightarrow 0$ calculated with rotationally adiabatic capture theory \citep{stoecklin1991rate,clary1993rate}.  
However, the maximum value in the rate coefficients measured in this work occurred near $T_{max}$ = 100 K (for CH $^{2}\Pi$ + CH$_{2}$O $^{1}A_{1}$) while the maximum calculated for another somewhat electronically similar doublet+singlet system (OH $^{2}\Pi$ + HBr $^{1}\Sigma$) occurred at $T_{max}$ = 20 K.  
This difference in $T_{max}$ is likely due to differences in the electronic effects in the long-range potential between CH + CH$_{2}$O \citep{clary1993rate}.  
\textnormal{Additionally, the maximum value of $k_{1}(T)$ is somewhat uncertain due to the uncertainty of the experimental measurements and lack of experimental measurements between 133 - 298 K.  }
Values for parameters of the best fit functions shown in Figure \ref{fig:k1(T)_and_All_Fits_vs_T} are given in Table \ref{tab:k_Mod_Arrhen_Fit_Parameters}.  
Furthermore, values for parameters from Table \ref{tab:k_Mod_Arrhen_Fit_Parameters} were converted from fits where $\tau = 1$ to $\tau = 300$ using Equation \ref{eqn:Convert_ModArrh} and are given in Table \ref{tab:k_Mod_Arrhen_Fit_Parameters_tau_300}.  

    \begin{deluxetable*}{cccccc}%k_Mod_Arrhen_Fit_Parameters
    \tablecaption{
        Modified Arrhenius equation, Equation \ref{eqn:ModArrhClassic}, parameters from fits to experimental values of $k_{1}(T)$
        \label{tab:k_Mod_Arrhen_Fit_Parameters}
    }
    \tablehead{
        %%%%%%%%%%%%%%%%%%%%%%%%%%%%%%%%%%%%%%%%%%%%%%%%%%
        \colhead{
            Temperature Range
        } & 
        \colhead{
            Curve in Figure \ref{fig:k1(T)_and_All_Fits_vs_T}
        } & 
        \colhead{
            $A$
        } &
        \colhead{
            $\tau$
        } &
        \colhead{
            $n$
        } &
        \colhead{
            $E_{a}/k_{B}$
        } \\
        %%%%%%%%%%%%%%%%%%%%%%%%%%%%%%%%%%%%%%%%%%%%%%%%%%
        \colhead{
            (K)
        } & 
        \colhead{} & 
        \colhead{
            ($cm^{3}s^{-1}$)
        } &
        \colhead{
            ($K$)
        } &
        \colhead{
            (Unitless)
        } &
        \colhead{
            ($K$)
        }
        %%%%%%%%%%%%%%%%%%%%%%%%%%%%%%%%%%%%%%%%%%%%%%%%%%
    }
    %\colnumbers
    \startdata
        %%%%%%%%%%%%%%%%%%%%%%%%%%%%%%%%%%%%%%%%%%%%%%%%%%
        31 - 670 & 
        Brown dotted curve [1] & 
        $(7.68 \pm 4.95)\times 10^{-7}$ &
        1\tablenotemark{a} &
        $-1.26 \pm 0.11$ & 
        $91.83 \pm 10.66$ \\
        %%%%%%%%%%%%%%%%%%%%%%%%%%%%%%%%%%%%%%%%%%%%%%%%%%
        298 - 670 & 
        Black curve [2] & 
        $(1.57 \pm 0.14)\times 10^{-10}$ &
        1\tablenotemark{a} &
        0\tablenotemark{a} & 
        $-260 \pm 30$ \\
        %%%%%%%%%%%%%%%%%%%%%%%%%%%%%%%%%%%%%%%%%%%%%%%%%%
        300 - 3000 & 
        Blue long dashed curve [3] & 
        $7.62\times 10^{-10}$ &
        1\tablenotemark{a} &
        $-0.32$ & 
        $-386$ \\
        %%%%%%%%%%%%%%%%%%%%%%%%%%%%%%%%%%%%%%%%%%%%%%%%%%
        31 - 133 & 
        Orange dot-dot-dash curve [1] & 
        $(4.20 \pm 0.23)\times 10^{-10}$ &
        1\tablenotemark{a} &
        %$0.1\Bar{6}$\tablenotemark{a} & 
        $1/6$\tablenotemark{a} & 
        0\tablenotemark{a} \\
        %%%%%%%%%%%%%%%%%%%%%%%%%%%%%%%%%%%%%%%%%%%%%%%%%%
        31 - 133 & 
        Yellow short dashed curve [1] & 
        $(2.15 \pm 1.03)\times 10^{-10}$ &
        1\tablenotemark{a} &
        $0.32 \pm 0.11$ & 
        0\tablenotemark{a} \\
        %%%%%%%%%%%%%%%%%%%%%%%%%%%%%%%%%%%%%%%%%%%%%%%%%%
    \enddata
    \tablerefs{
        [1]This work; 
        [2]\cite{nguyen2014theoretical};
        [3]\cite{zabarnick1988temperature}; 
    }
    \tablenotetext{a}{
        Value held constant during fit.  
    }
    \end{deluxetable*}

    \begin{deluxetable*}{cccccc}%k_Mod_Arrhen_Fit_Parameters
    \tablecaption{
        Parameters from Table \ref{tab:k_Mod_Arrhen_Fit_Parameters} converted to modified Arrhenius, Equation \ref{eqn:ModArrhClassic}, fits where $\tau = 300$ using Equation \ref{eqn:Convert_ModArrh}.  
        \label{tab:k_Mod_Arrhen_Fit_Parameters_tau_300}
    }
    \tablehead{
        %%%%%%%%%%%%%%%%%%%%%%%%%%%%%%%%%%%%%%%%%%%%%%%%%%
        \colhead{
            Temperature Range
        } & 
        \colhead{
            Curve in Figure \ref{fig:k1(T)_and_All_Fits_vs_T}
        } & 
        \colhead{
            $A$
        } &
        \colhead{
            $\tau$
        } &
        \colhead{
            $n$
        } &
        \colhead{
            $E_{a}/k_{B}$
        } \\
        %%%%%%%%%%%%%%%%%%%%%%%%%%%%%%%%%%%%%%%%%%%%%%%%%%
        \colhead{
            (K)
        } & 
        \colhead{} & 
        \colhead{
            ($cm^{3}s^{-1}$)
        } &
        \colhead{
            ($K$)
        } &
        \colhead{
            (Unitless)
        } &
        \colhead{
            ($K$)
        }
        %%%%%%%%%%%%%%%%%%%%%%%%%%%%%%%%%%%%%%%%%%%%%%%%%%
    }
    %\colnumbers
    \startdata
        %%%%%%%%%%%%%%%%%%%%%%%%%%%%%%%%%%%%%%%%%%%%%%%%%%
        31 - 670 & 
        Brown dotted curve [1] &
        $5.81\times 10^{-10}$ &
        300\tablenotemark{a} &
        -1.26 & 
        91.83 \\ 
        %%%%%%%%%%%%%%%%%%%%%%%%%%%%%%%%%%%%%%%%%%%%%%%%%%
        298 - 670 & 
        Black curve [2] & 
        $1.57\times 10^{-10}$ &
        300\tablenotemark{a} &
        0\tablenotemark{a} & 
        -260 \\
        %%%%%%%%%%%%%%%%%%%%%%%%%%%%%%%%%%%%%%%%%%%%%%%%%%
        300 - 3000 & 
        Blue long dashed curve [3] & 
        $1.23\times 10^{-10}$ & %$7.62\times 10^{-10}$
        300\tablenotemark{a} &
        $-0.32$ & 
        $-386$ \\
        %%%%%%%%%%%%%%%%%%%%%%%%%%%%%%%%%%%%%%%%%%%%%%%%%%
        31 - 133 & 
        Orange dot-dot-dash curve [1] &  
        $1.09\times 10^{-9}$ & %$(4.20 \pm 0.23)\times 10^{-10}$
        300\tablenotemark{a} &
        $1/6$\tablenotemark{a} & 
        %$0.1\Bar{6}$\tablenotemark{a} & 
        0\tablenotemark{a} \\
        %%%%%%%%%%%%%%%%%%%%%%%%%%%%%%%%%%%%%%%%%%%%%%%%%%
        31 - 133 & 
        Yellow short dashed curve [1] &
        $1.33\times 10^{-9}$ & %$(2.15 \pm 1.03)\times 10^{-10}$
        300\tablenotemark{a} &
        0.32 & 
        0\tablenotemark{a} \\
        %%%%%%%%%%%%%%%%%%%%%%%%%%%%%%%%%%%%%%%%%%%%%%%%%%
    \enddata
    \tablerefs{
        [1]This work; 
        [2]\cite{nguyen2014theoretical};
        [3]\cite{zabarnick1988temperature}; 
    }
    \tablenotetext{a}{
        Value held constant during fit.  
    }
    \end{deluxetable*}

\FloatBarrier
%%%%%%%%%%%%%%%%%%%%%%%%%%%%%%%%%%%%%%
%%IMPACT ON CSM/ISM MODELS
%%%%%%%%%%%%%%%%%%%%%%%%%%%%%%%%%%%%%%

\section{Impact on Calculated Abundances in Astrophysical Environments} \label{sec:ModelImpact}

The newly measured and theoretically calculated rate coefficients were used to investigate the impact of the CH + CH$_{2}$O reaction on the abundances of CH$_2$O, H$_2$CCO, and HCO \textnormal{and }compared \textnormal{these} to the calculated abundances without the new coefficients, using the UMIST Rate12 database \citep{mce13}.  
The detailed parameters used in our calculations are given in Table \ref{tab:Arr}, where `Rate12' refers to the UMIST Database which includes only the HCO + CH$_2$ channel (Reaction \ref{eqn:CH_CH2O-CH2_HCO}), `CT' to the T$^{1/6}$-dependent rate coefficient based on capture theory, and `MA' to the modified Arrhenius fit to the experimental data over the 31--670 K range (see Figure \ref{fig:k1(T)_and_All_Fits_vs_T}).  

\begin{table*}
    \caption{
        Parameters for Equation \ref{eqn:ModArrhAstro} used in calculating temperature-dependent rate coefficients for the CH + CH$_2$O reaction.
    }
    \centering
    \begin{tabular}{ccccccccccc}
    \hline \hline
    \noalign{\smallskip}
    Reaction & Products & \multicolumn {3}{c}{Rate12} & \multicolumn{3}{c}{CT} & \multicolumn{3}{c}{MA} \\
         & & {$\alpha$} & {$\beta$} & {$\gamma$}  & {$\alpha$} & {$\beta$} & {$\gamma$}  & {$\alpha$} & {$\beta$} & {$\gamma$} \\
    \noalign{\smallskip}
    \hline
    \ref{eqn:CH_CH2O-CH2_HCO} & HCO + CH$_2$ & 9.21 $\times$ 10$^{-12}$ & 0.70 & 2000 & 2.18 $\times$ 10$^{-11}$ & 0.17 & 0.0 & 1.11 $\times$ 10$^{-11}$ & -1.26 & 91.83 \\
    \ref{eqn:CH_CH2O-H_H2CCO} & H$_2$CCO + H & - & - & - &8.94 $\times$ 10$^{-10}$ & 0.17 & 0.0 & 4.76 $\times$ 10$^{-10}$ & -1.26 & 91.83  \\
    \ref{eqn:CH_CH2O-CH3_CO} & CH$_3$ + CO & - & - & - & 1.74 $\times$ 10$^{-10}$ & 0.17 & 0.0 & 9.30 $\times$ 10$^{-11}$ & -1.26 & 91.83 \\
    \hline
    \end{tabular}
    \tablecomments{
        Parameters for models based on capture theory (CT) and modified Arrhenius (MA) fits were calculated using fit curve parameters given in Table \ref{tab:k_Mod_Arrhen_Fit_Parameters_tau_300} where the alpha value for each channel was scaled according to the theoretical 300 K yields of Reactions \ref{eqn:CH_CH2O-H_H2CCO}-\ref{eqn:CH_CH2O-CH2_HCO}.  
    }
    \label{tab:Arr}
\end{table*}

%%%%%%%%%%%%%%%%%%%%%%%%%%%%%%%%%%%%%%
%%ISC MODELING
%%%%%%%%%%%%%%%%%%%%%%%%%%%%%%%%%%%%%%
\subsection{Dark Interstellar Clouds}

We have investigated the implications of our newly measured rate coefficients in a model of a dark interstellar cloud \textnormal{with a visual extinction of 10 mags}. 
We have modeled the gas-phase chemistry of two separate densities, n(H$_2$) = 10$^4$ and 10$^5$ cm$^{-3}$, each at temperatures of 10, 20 and 30 K, and for each of the three parameterizations of the rate coefficient given in Table \ref{tab:Arr}.  \textnormal{We used the low-metal elemental abundances believed to be appropriate for dark clouds \citep{mce13}.} 

We have looked, in particular, at the abundances of CH and CH$_2$O as well as the main astronomically observable products of the reaction, CO, HCO, CH$_3$ and H$_2$CCO. 
Our calculations show that the abundances of the reactants CH and CH$_2$O are unchanged in the models as Reaction \ref{eqn:Rxn_CH_CH2O-Prod} represents only a minor loss of these species at all times, densities and temperatures. 
This is not surprising as CH is destroyed rapidly in reaction with species more abundant than formaldehyde, in particular the atoms O, N and H.  
Formaldehyde is predominantly destroyed by fast proton transfer reactions followed by dissociative recombination with electrons.

\begin{figure}[H] %Dark_Cloud_Abundances_vs_t
    \plottwo
        %{figures/Dark_Cloud_Abundances_vs_t.png}
        {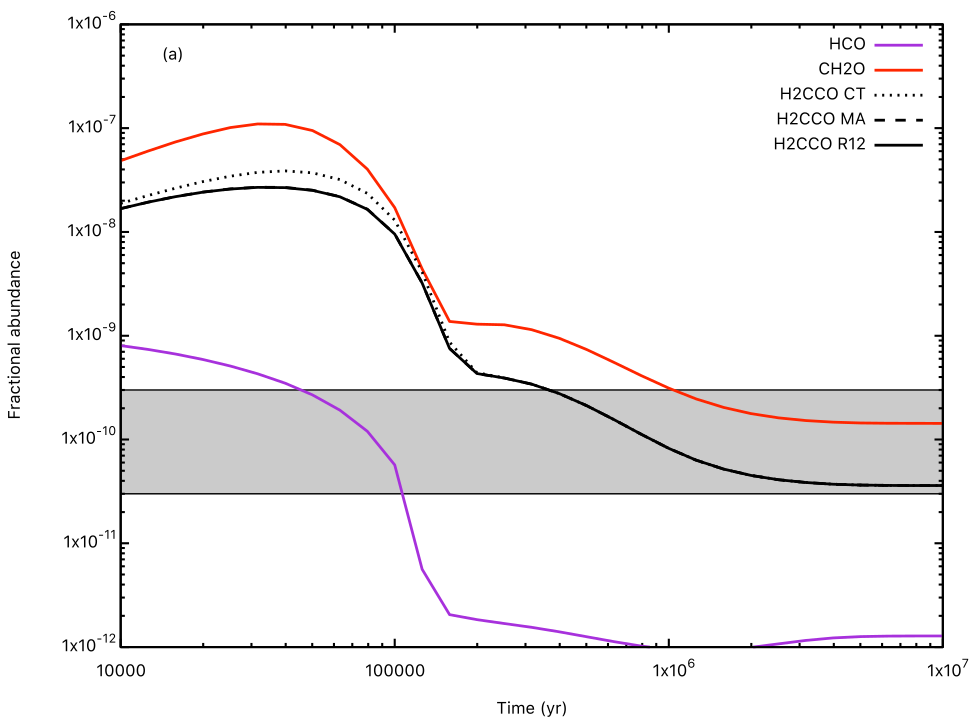}
        {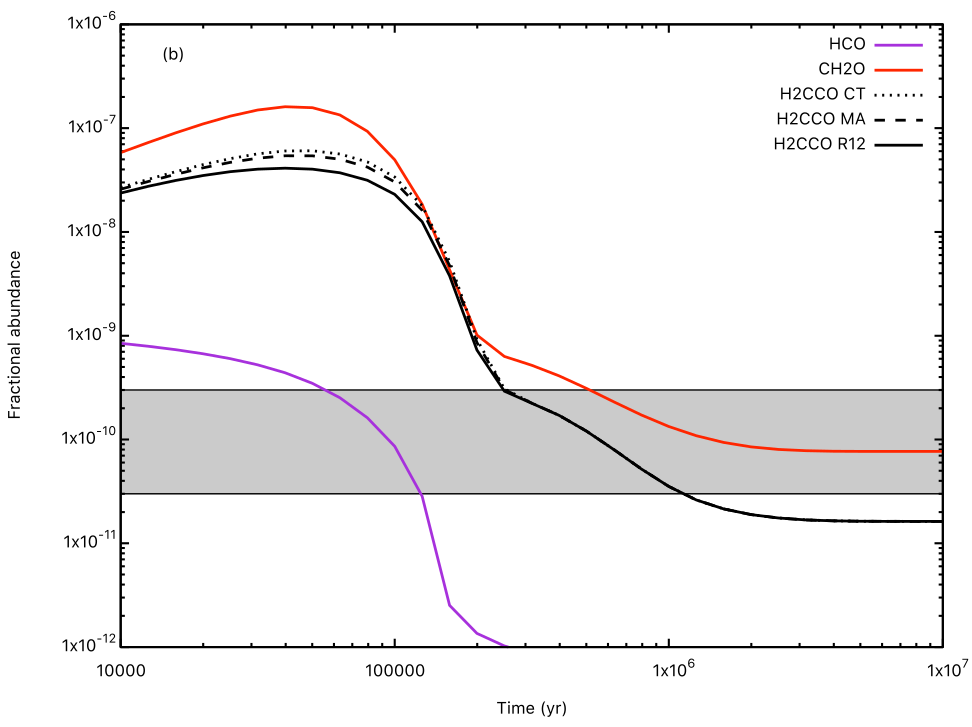}
    \caption{
        Abundances versus time of key reagents and products of the title reaction in cloud models with n(H$_2$) = 10$^5$ cm$^{-3}$ and A$_V$ = 10 mags for (a) T = 10 K and (b) T = 30 K. The abundances of HCO and CH$_2$O are identical in the CT, MA and R12 calculations. The grey box indicates the range in ketene abundances measured in dark clouds by \citet{rui07} and \citet{agu10}.
        \label{fig:dark}
    }
\end{figure}

Similarly, the products of the title reaction have abundances which do not significantly differ from those in the Rate12 model, with the exception of H$_2$CCO which
shows a small increase in abundance at early times, that is less than 2 $\times$ 10$^5$ yr, at all temperatures for n(H$_2$) = 10$^5$ cm$^{-3}$ shown in Figure \ref{fig:dark}. 
In the UMIST database, the products of the title reaction are assumed to be CH$_2$ and HCO with an activation energy
barrier of 2000 K \citep{mit84}, taken from the compilation by \cite{wes80}.

\citet{nguyen2014theoretical} determined that the major molecular product in the title reaction was ketene which thus provides a new route when compared to the UMIST Rate12 database.  
The impact of this new route is seen only in the higher density models at early times (Figure \ref{fig:dark}) where we find that between 8 and 21\% of ketene is formed by this reaction. 
The largest contributions, 16--21\%, to ketene formation occur at T = 30 K.  
In all cases, the percentage contribution is larger for the CT rate coefficient than for the MA fit to the laboratory data.  Additionally, at 10 K the MA model yields unnoticeable differences to the H$_{2}$CCO abundance while the CT model yields a similar contribution to the H$_{2}$CCO abundance as in the 30 K model.  This is due to the divergence of the MA and CT fit curves below 30 K.  
The increases in the ketene fractional abundance are relatively modest, however, with its maximum abundance, which occurs at 4--5 $\times$ 10$^4$ yr, varying as 4$\times$10$^{-8}$, 5$\times$10$^{-8}$, and 6$\times$10$^{-8}$ in the Rate12, MA, and CT models, respectively at 30 K and a change of a factor of $\sim$2 is observed in ketene at 10 K between CT and Rate12 models.  
If we use the smaller rate coefficient, 5 $\times$ 10$^{-11}$ cm$^3$ s$^{-1}$, recommended by \citet{bau05} for the O + C$_2$H$_3$ reaction in the range 250--2000 K (Section \ref{sec:intro}), then the maximum ketene fractional abundance is reduced slightly to 3--5 $\times$ 10$^{-8}$.

Finally, noting that \citet{nguyen2014theoretical} suggested that the ketene yield in reaction \ref{eqn:CH_CH2O-H_H2CCO} should increase with decreasing temperature, we ran a series of calculations with a 100\%, rather than 82\% yield for this channel. As expected, the increase in the H$_2$CCO abundance was minimal, limited to around 5--6\% for the models presented here. 

\FloatBarrier
%%%%%%%%%%%%%%%%%%%%%%%%%%%%%%%%%%%%%%
%%CSE MODELING
%%%%%%%%%%%%%%%%%%%%%%%%%%%%%%%%%%%%%%
\subsection{Circumstellar Envelopes}

We have also investigated the implications of the parameterized rate coefficients listed in Table \ref{tab:Arr} on the chemistry within the CSEs of AGB stars.  
Our model is based on the publically available UMIST Database for Astrochemistry CSE model \citep{mce13}\footnote{\url{http://udfa.ajmarkwick.net/index.php?mode=downloads}}, where we changed the assumed gas temperature structure to a power-law,
\begin{equation}
    T(r) = T_* \left( \frac{R_*}{r} \right)^\epsilon,
\end{equation}
with $T_*$ and $R_*$ the stellar temperature and radius, and $\epsilon$ the exponent characterising the power-law \citep{vds18a}. 
We calculated a grid of models, where we varied over the mass-loss rate of outflow, $\dot{M} = 10^{-5}$ and $10^{-7}$ M$_\odot$ yr$^{-1}$, the stellar temperature $T_* = 2000$ and 2300 K, the power-law exponent $\epsilon = 0.5$ and 0.7. 
We assume that stellar radius $R_*$ = $5 \times 10^{13}$ cm and a constant expansion velocity of 15 km s$^{-1}$.  
Both an O-rich and a C-rich outflow are investigated. 
The parent species and their initial abundances are taken from \citet{agu10}.

The results are similar to those from the dark cloud models.
We find that the inclusion of the CT and MA rate coefficients \textnormal{does} not noticeably affect the abundance profiles of CH and CH$_2$O or any of the observable reaction products.
The largest changes are seen for C-rich outflows because of the higher density of the reactants.  
For both C-rich and O-rich outflows, higher density outflows with a colder temperature structure, i.e. $T_* = 2000$ K and $\epsilon = 0.5$, result in larger changes. 
The column densities of the species considered change by maximally 2\% relative to the Rate12 reaction rate coefficients in both the O-rich and C-rich CSE. 
These changes are too small to be observable.  
The largest change of $\sim$2\% corresponds to the decrease in column density of CH$_{2}$O.
The column density of H$_2$CCO increases by maximally 1\%. 
The additional route to form ketene hence also increases its abundance in CSEs, albeit only very slightly due to the difference in physical structure of the outflow.
\textnormal{Using a 100\% yield for reaction channel \ref{eqn:CH_CH2O-H_H2CCO} instead of 82\% does not result in a larger increase of the ketene abundance.}
The column densities of the other products increase by less than 1\%.

\FloatBarrier
%%%%%%%%%%%%%%%%%%%%%%%%%%%%%%%%%%%%%%
%%CONCLUSIONS AND PROSPECTS
%%%%%%%%%%%%%%%%%%%%%%%%%%%%%%%%%%%%%%
\section{Conclusions and Prospects} \label{sec:Conclusions}
Over the temperature range 31 - 133 K, rate coefficients for the reaction of CH + CH$_{2}$O, $k_{1}(T)$, have been determined to be very large such that the reaction is at the collision limit.  
Values of $k_{1}(T)$ near $\sim$100 K are among the largest measured rate coefficient values for a neutral-neutral gas phase reaction below 300 K (within the largest 4 of the 620 listed in the UMIST RATE12 database).  
$k_{1}(T)$ at 30 K is among the 10 largest rate coefficient values for a neutral-neutral gas phase reaction included in the UMIST RATE12 database (many of which are extrapolations from measurements of C atom reactions at 300 K).  
Below $\sim$70 K, measured values of $k_{1}(T)$ were observed to decrease with a decrease in temperature.  
This positive temperature dependence is not predicted by classical capture rate theory but is predicted by the more detailed adiabatic capture rate theory.  
The values of $k_{1}(T)$ have been parameterized both using a simple modified Arrhenius equation as well as using an $A\times T^{1/6}$ fit based on low temperature AC theory.  
These parameterizations were then added to the UMIST Rate12 astrochemical \textnormal{chemical network} for two model scenarios: dark interstellar clouds and circumstellar envelopes.  
The change in CH$_{2}$O abundance in these two environments was essentially unchanged with our inputted $k_{1}(T)$, since formaldehyde is destroyed rapidly in reactions with ions and with other more abundant atoms and radicals than CH.  
However, the dominant molecular product from CH + CH$_{2}$O is ketene, H$_{2}$CCO, and significant abundance changes in ketene were observed in some of the model runs. 
The observation of a large rate coefficient decreasing with a decrease in temperature at temperatures relevant for interstellar space, as was measured for the reaction CH + CH$_{2}$O, might be more general than is presently acknowledged. 
Our extrapolation of $k_{1}(T)$ to $T$ below our measured values has a large uncertainty, but this uncertainty would be significantly reduced if there was better theoretical understanding of collision rate theory at low $T$.  
Although AC theory and $\mu$j-VTST have been shown to calculate qualitatively correct temperature dependencies to within a factor of two for measured reaction rate coefficients of some systems at the collision limit, even better theoretical understanding of collision rate theory at low $T$ would allow for higher confidence in extrapolation of fits of measured rate coefficients to lower temperatures.  
\textnormal{
    Until more experimental or theoretical results are available to extend the temperature range over which the rate coefficient is determined, we recommend that the MA fit (brown dotted curve in Figure \ref{fig:k1(T)_and_All_Fits_vs_T}) is applied over the range of 38.5 $\leq T$ (K) $\leq$ 670 and the $AT^{n}$ fit (yellow short dashed curve in Figure \ref{fig:k1(T)_and_All_Fits_vs_T}) is applied over the range of 0 $\leq T$ (K) $\leq$ 38.5, if an extrapolation is to be made below 31 K, for example for use in astrochemical simulations.  
    We recommend that these two parameterized fits be input, over their respective temperature ranges, into reaction databases such as UMIST Rate12 and the KIDA.  
}

\acknowledgments

%%%%%%%%%%%%%%%%%%%%%%%%%%%%%%%%%%%%%%
%%ACKNOWLEDGEMENTS
%%%%%%%%%%%%%%%%%%%%%%%%%%%%%%%%%%%%%%
\section{Acknowledgements} \label{sec:Acknowledgements}
This project has received funding from the European Research Council (ERC) under the European Union’s Horizon 2020 research and innovation programme (grant agreement No 646758).  
We would like to thank the mechanical and electronics workshops in the School of Chemistry at the University of Leeds for support.  
We would like to thank John Plane for helpful discussions on classical capture theory.  
TJM is grateful to the STFC for support through grant ST/P000321/1 and to the Institute for Theory and Computation for hospitality.
MVdS acknowledges support from the Research Foundation Flanders (FWO) through grant 12X6419N.
We would also like to thank two anonymous referees for their prompt response and insightful comments.

\software{\textnormal{LIFBASE \citep{luque1999lifbase}}}
\software{\textnormal{UDfA \citep{mce13}}}

\FloatBarrier
%%%%%%%%%%%%%%%%%%%%%%%%%%%%%%%%%%%%%%
%%APPENDIX
%%%%%%%%%%%%%%%%%%%%%%%%%%%%%%%%%%%%%%
\appendix
\section{UV Absorption} \label{sec:Appendix_UVAbs}

\begin{figure}[ht] %UV_Abs_CH2O
    \plotone{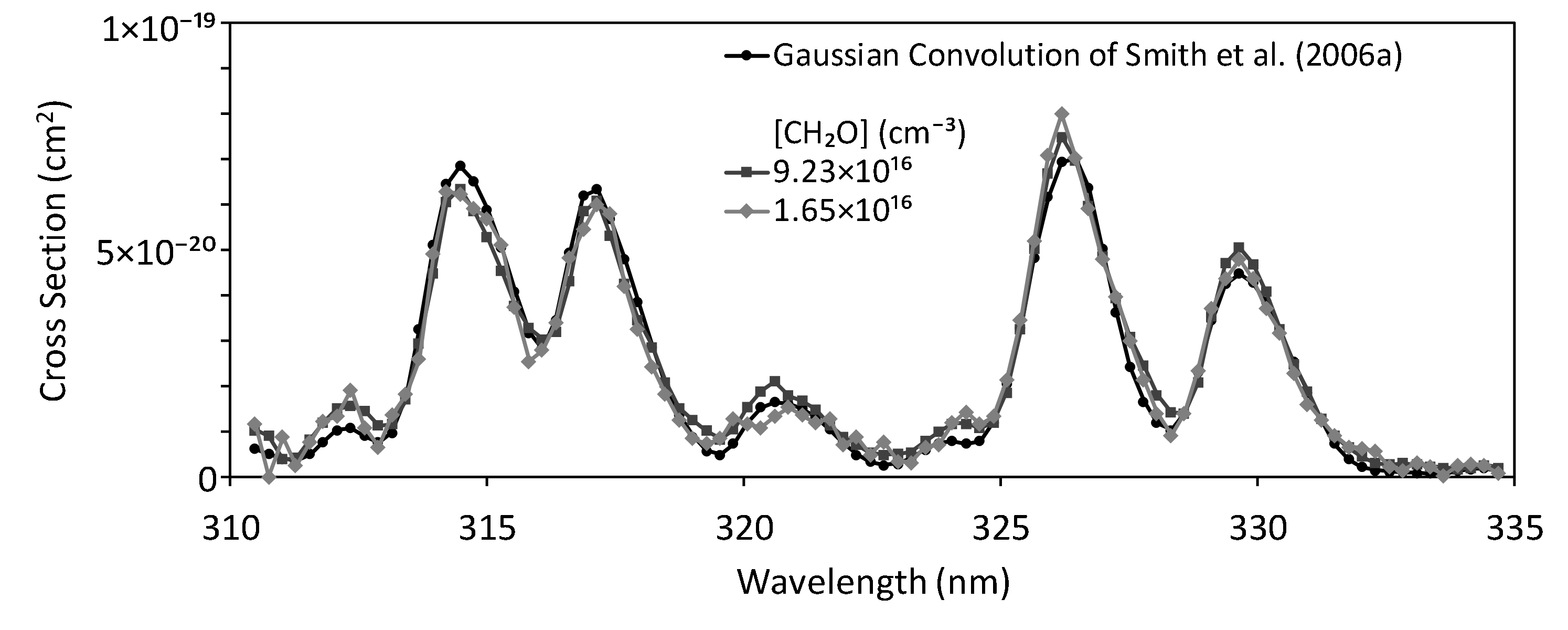}
    \caption{
        Absorption spectra of formaldehyde at various concentrations of CH$_{2}$O fitted to the convolution of a Gaussian function of  0.75 nm FWHM with a 0.0035 nm high resolution spectrum from \cite{smith2006absorption}.  
        \label{fig:UV_Abs_CH2O}
    }
\end{figure}

UV absorption spectra of formaldehyde gas were utilized in order to determine the concentration of formaldehyde in each final mixture of gas in low temperature kinetics measurements.  
Two representative UV absorption spectra are shown in Figure \ref{fig:UV_Abs_CH2O}.  
In order to fit UV absorption spectra collected in this study, a least-squares minimization analysis was performed comparing the collected spectra to a modified version of a literature UV absorption spectrum.  
First, a high-resolution UV absorption spectrum from \cite{smith2006absorption} was convoluted with a 0.75 nm Gaussian function in order to match the resolution of the spectrometer in this study.  
The convoluted spectrum was then linearly interpolated onto the wavelength grid of the spectra collected in this study.  
An initial guess of the number density of formaldehyde in the absorption cell, $N^{AbsCell}_{CH_{2}O}$ (cm$^{-3}$), was then utilized to convert absorbance, $A$, to absorption cross section, $\sigma$.   
A least-squares minimization analysis was then performed by varying the estimated $N^{AbsCell}_{CH_{2}O}$ in the data spectra in order to obtain a best fit of $N^{AbsCell}_{CH_{2}O}$, while the total pressure measured in the absorption cell was then utilized to calculate the total number density, $N^{AbsCell}_{total}$ (cm$^{-3}$).  
$N^{AbsCell}_{CH_{2}O}$ was then divided by $N^{AbsCell}_{total}$ in order to calculate the fraction of formaldehyde gas in the cell, and the value then \textnormal{adopted} as the fraction of formaldehyde in the low temperature flows generated by the Laval nozzles.  
The statistical error in the fitted  $N^{AbsCell}_{CH_{2}O}$ values were determined to be $\lesssim$2\% by taking the standard error of the slope of $A/l^{AbsCell}$ versus $\sigma^{lit}$ where $l^{AbsCell}$ is the path length, and $A$ and $\sigma^{lit}$ are the wavelength dependent absorbance values and literature cross section values from \cite{smith2006absorption} respectively.  
However, for [CH$_{2}$O] values used in second order plots, Figure \ref{fig:Int_Sub_Second_Order_Plots}, the uncertainty was dominated by the $\sim$10\% uncertainty of the density of the cold flows.  

\FloatBarrier
\section{Second Order Plots} \label{sec:Second_Order_Plots}

\begin{figure}[ht] %Not_Subtracted_Second_Order_Plots
    \plottwo
        {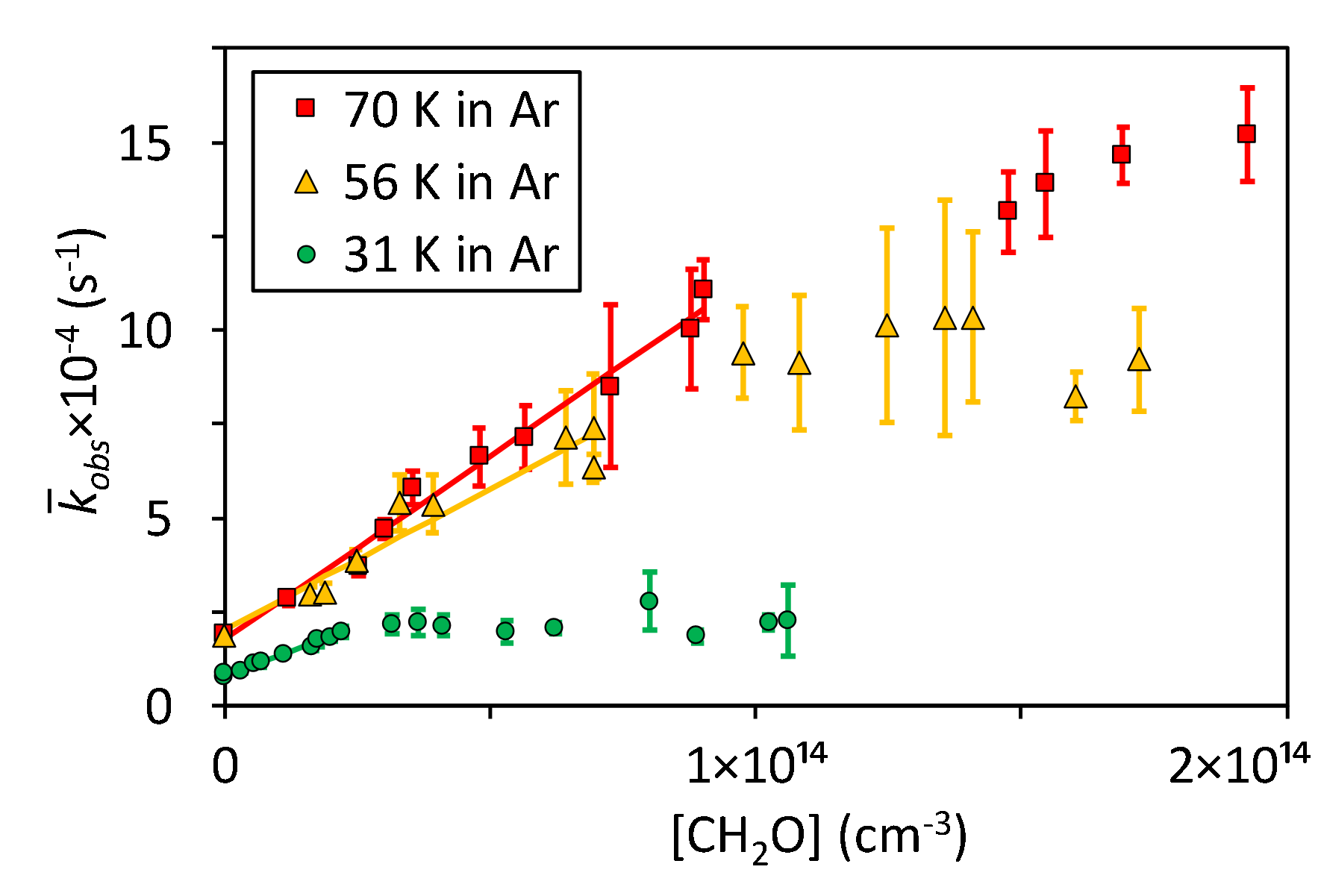}
        {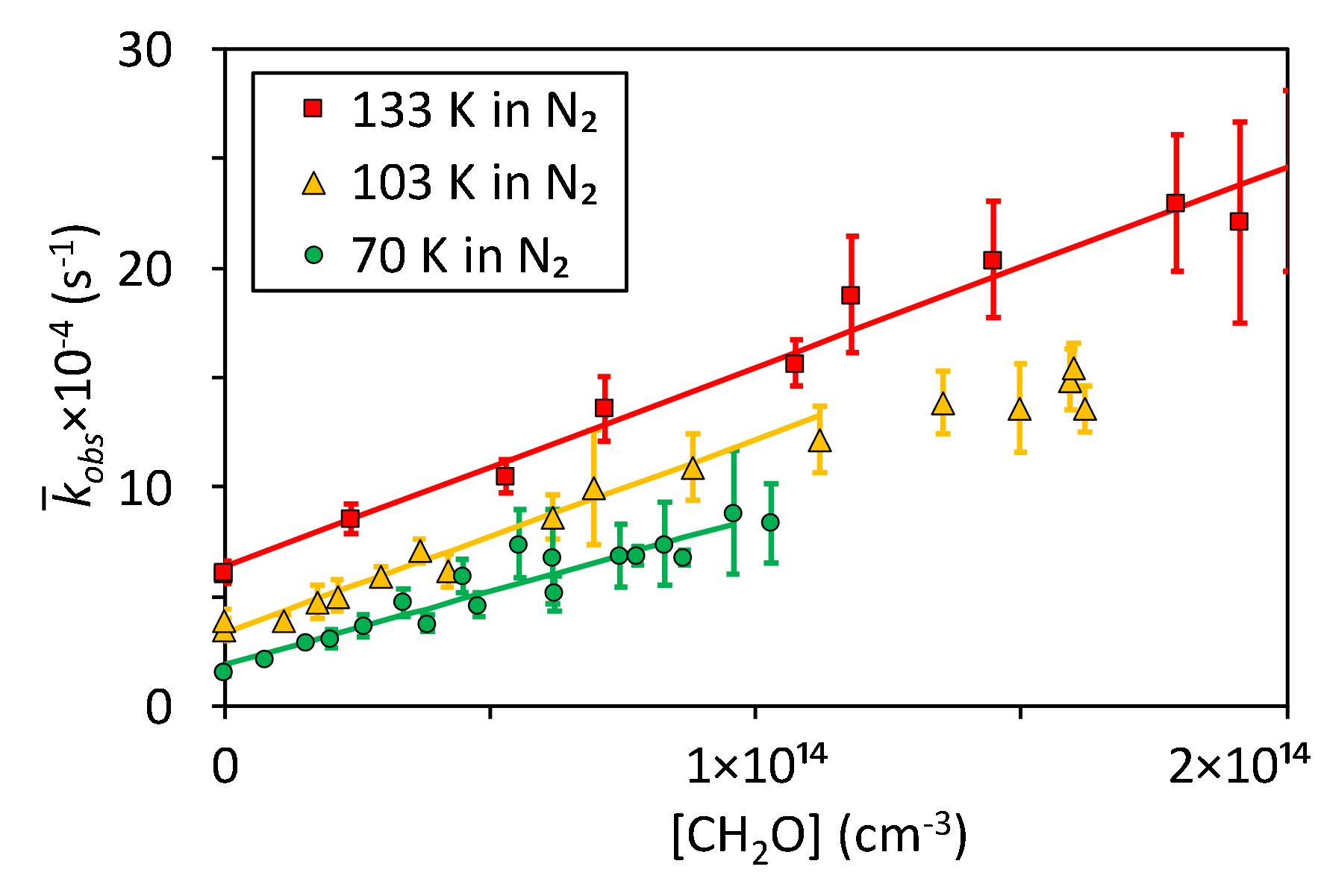}
    \caption{
        Average rate of loss of CH versus the concentration of formaldehyde at various temperatures along with linear fits at each temperature.  
        Error bars represent one standard deviation of fits of $k_{obs}$ from at least 5 experimental decay traces. 
        \label{fig:Nonsub_Second_Order_Plots}}
\end{figure}

Un-subtracted plots of the average value of fits of the rate of loss of CH ($\bar{k}_{obs}$) versus [CH$_{2}$O] are shown in Figure \ref{fig:Nonsub_Second_Order_Plots}.  
The \textnormal{fitted $k_{int}$ values from linear fits} represent loss of CH due to diffusion as well as reaction with species besides CH$_{2}$O \textnormal{(primarily N$_{2}$ when present and, with probable minor contributions from the precursor CHBr$_{3}$, one and two-photon photolysis products of CHBr$_{3}$, and as stated by the manufacturer, the CHBr$_{3}$ stabilizer 2-methyl-2-butene, which was present at 60-120 ppm in the CHBr$_{3}$ liquid)}.  
Therefore, once differences in the rates of CH diffusion (due to differences in flow temperature and density) are subtracted from each \textnormal{$k_{int}$}, the remaining contribution to \textnormal{$k_{int}$} is primarily accounted for by the reaction of CH with N$_{2}$ (when N$_{2}$ was present in the flow field) to within the uncertainty of our \textnormal{$k_{int}$} values when compared with low temperature rate coefficients for the reaction of CH + N$_{2}$ measured by \cite{rowe1998determination}.  
\textnormal{The rate coefficients for three-body association of CH + N$_{2}$  from \cite{rowe1998determination} down to 53 K were of similar magnitude but somewhat smaller than those obtained through analysis of $k_{int}$ values in this work, however, there is a relatively high uncertainty in obtaining a rate coefficient for CH + N$_{2}$ from analysis of $k_{int}$ values.}  

\FloatBarrier
%%%%%%%%%%%%%%%%%%%%%%%%%%%%%%%%%%%%%%
%%REFERENCES
%%%%%%%%%%%%%%%%%%%%%%%%%%%%%%%%%%%%%%
\bibliography{CH_CH2O_Manuscript}

\begin{thebibliography}{}
\expandafter\ifx\csname natexlab\endcsname\relax\def\natexlab#1{#1}\fi
\providecommand{\url}[1]{\href{#1}{#1}}

\bibitem[{{Ag{\'u}ndez} {et~al.}(2010){Ag{\'u}ndez}, {Cernicharo}, \&
  {Gu{\'e}lin}}]{agu10}
{Ag{\'u}ndez}, M., {Cernicharo}, J., \& {Gu{\'e}lin}, M. 2010, \apjl, 724, L133

\bibitem[{Anderson {et~al.}(1996)Anderson, Lanning, Barrell, Miyagishima,
  Jones, \& Wolfe}]{anderson1996sources}
Anderson, L.~G., Lanning, J.~A., Barrell, R., {et~al.} 1996, Atmos. Env., 30,
  2113

\bibitem[{Atreya(2010)}]{atreya2010significance}
Atreya, S.~K. 2010, Far. Dis., 147, 9

\bibitem[{{Baulch} {et~al.}(2005){Baulch}, {Bowman}, {Cobos}, {Cox}, {Just},
  {Kerr}, {Pilling}, {Stocker}, {Troe}, {Tsang}, {Walker}, \&
  {Warnatz}}]{bau05}
{Baulch}, D.~L., {Bowman}, C.~T., {Cobos}, C.~J., {et~al.} 2005, Journal of
  Physical and Chemical Reference Data, 34, 757

\bibitem[{Canosa {et~al.}(1997)Canosa, Sims, Travers, Smith, \&
  Rowe}]{canosa1997reactions}
Canosa, A., Sims, I., Travers, D., Smith, I., \& Rowe, B. 1997, \aap, 323, 644

\bibitem[{Caravan {et~al.}(2014)Caravan, Shannon, Lewis, Blitz, \&
  Heard}]{caravan2014measurements}
Caravan, R.~L., Shannon, R.~J., Lewis, T., Blitz, M.~A., \& Heard, D.~E. 2014,
  JCPA, 119, 7130

\bibitem[{Carlier {et~al.}(1986)Carlier, Hannachi, \&
  Mouvier}]{carlier1986chemistry}
Carlier, P., Hannachi, H., \& Mouvier, G. 1986, Atmos. Env. (1967), 20, 2079

\bibitem[{Clary(1994)}]{clary1994rate}
Clary, D. 1994, AIP Conference Proceedings, 312, 405

\bibitem[{Clary {et~al.}(1993)Clary, Stoecklin, \& Wickham}]{clary1993rate}
Clary, D.~C., Stoecklin, T.~S., \& Wickham, A.~G. 1993, Far. Trans., 89, 2185

\bibitem[{Cooke \& Sims(2019)}]{cooke2019experimental}
Cooke, I., \& Sims, I.~R. 2019, ACS Earth and Space Chemistry

\bibitem[{{Danks} {et~al.}(1984){Danks}, {Federman}, \& {Lambert}}]{danks84}
{Danks}, A.~C., {Federman}, S.~R., \& {Lambert}, D.~L. 1984, \aap, 130, 62

\bibitem[{Fenimore(1971)}]{fenimore1971formation}
Fenimore, C. 1971, Symposium (International) on Combustion, 13, 373

\bibitem[{Georgievskii \& Klippenstein(2005)}]{georgievskii2005long}
Georgievskii, Y., \& Klippenstein, S.~J. 2005, \jcp, 122, 194103

\bibitem[{G{\'o}mez~Mart{\'\i}n {et~al.}(2014)G{\'o}mez~Mart{\'\i}n, Caravan,
  Blitz, Heard, \& Plane}]{gomez2014low}
G{\'o}mez~Mart{\'\i}n, J., Caravan, R., Blitz, M., Heard, D., \& Plane, J.
  2014, JCPA, 118, 2693

\bibitem[{Goulay {et~al.}(2009)Goulay, Trevitt, Meloni, Selby, Osborn, Taatjes,
  Vereecken, \& Leone}]{goulay2009cyclic}
Goulay, F., Trevitt, A.~J., Meloni, G., {et~al.} 2009, JAChS, 131, 993

\bibitem[{Grosjean {et~al.}(1993)Grosjean, Williams, \&
  Grosjean}]{grosjean1993ambient}
Grosjean, E., Williams, E.~L., \& Grosjean, D. 1993, Air \& Waste, 43, 469

\bibitem[{Heard(2018)}]{heard2018rapid}
Heard, D.~E. 2018, AcChR, 51, 2620

\bibitem[{Herzberg \& Johns(1969)}]{herzberg1969new}
Herzberg, G., \& Johns, J. 1969, \apj, 158, 399

\bibitem[{{Hirschfelder} {et~al.}(1964){Hirschfelder}, {Curtiss}, \&
  {Bird}}]{hirschfelder1964molecular}
{Hirschfelder}, J.~O., {Curtiss}, C.~F., \& {Bird}, R.~B. 1964, {Molecular
  theory of gases and liquids} (Wiley: New York)

\bibitem[{{Hudson} \& {Loeffler}(2013)}]{hud13}
{Hudson}, R.~L., \& {Loeffler}, M.~J. 2013, \apj, 773, 109

\bibitem[{Krasnopolsky(2009)}]{krasnopolsky2009photochemical}
Krasnopolsky, V.~A. 2009, Icarus, 201, 226

\bibitem[{Le~Picard {et~al.}(1998)Le~Picard, Canosa, Rowe, Brownsword, Smith,
  {et~al.}}]{rowe1998determination}
Le~Picard, S.~D., Canosa, A., Rowe, B.~R., {et~al.} 1998, Far. Trans., 94, 2889

\bibitem[{Lindner {et~al.}(1998)Lindner, Ermisch, \&
  Wilhelm}]{lindner1998multi}
Lindner, J., Ermisch, K., \& Wilhelm, R. 1998, Chemical physics, 238, 329

\bibitem[{Linstrom \& Mallard(2001)}]{linstrom2001nist}
Linstrom, P.~J., \& Mallard, W. 2001, NIST Chemistry webbook; NIST standard
  reference database No. 69,  National Institute of Standards and Technology.
\newblock \url{https://webbook.nist.gov/}

\bibitem[{Luque \& Crosley(1999)}]{luque1999lifbase}
Luque, J., \& Crosley, D.~R. 1999, SRI international report MP, 99

\bibitem[{{Maity} {et~al.}(2014){Maity}, {Kaiser}, \& {Jones}}]{mai14}
{Maity}, S., {Kaiser}, R.~I., \& {Jones}, B.~M. 2014, \apj, 789, 36

\bibitem[{Manohar \& Pal(2007)}]{manohar2007dipole}
Manohar, P.~U., \& Pal, S. 2007, CPL, 438, 321

\bibitem[{{Matthews} \& {Sears}(1986)}]{mat86}
{Matthews}, H.~E., \& {Sears}, T.~J. 1986, \apj, 300, 766

\bibitem[{{McElroy} {et~al.}(2013){McElroy}, {Walsh}, {Markwick}, {Cordiner},
  {Smith}, \& {Millar}}]{mce13}
{McElroy}, D., {Walsh}, C., {Markwick}, A.~J., {et~al.} 2013, \aap, 550, A36

\bibitem[{{McKellar}(1941)}]{mck41}
{McKellar}, A. 1941, Publ DAO, 7, 251

\bibitem[{Miller \& Bowman(1989)}]{miller1989mechanism}
Miller, J.~A., \& Bowman, C.~T. 1989, PrECS, 15, 287

\bibitem[{{Mitchell}(1984)}]{mit84}
{Mitchell}, G.~F. 1984, \apjs, 54, 81

\bibitem[{{Muller} {et~al.}(2014){Muller}, {Combes}, {Gu{\'e}lin}, {G{\'e}rin},
  {Aalto}, {Beelen}, {Black}, {Curran}, {Darling}, {V-Trung},
  {Garc{\'\i}a-Burillo}, {Henkel}, {Horellou}, {Mart{\'\i}n},
  {Mart{\'\i}-Vidal}, {Menten}, {Murphy}, {Ott}, {Wiklind}, \&
  {Zwaan}}]{muller14}
{Muller}, S., {Combes}, F., {Gu{\'e}lin}, M., {et~al.} 2014, \aap, 566, A112

\bibitem[{Nelson~Jr {et~al.}(1967)Nelson~Jr, Lide~Jr, \&
  Maryott}]{nelson1967selected}
Nelson~Jr, R.~D., Lide~Jr, D.~R., \& Maryott, A.~A. 1967, Selected values of
  electric dipole moments for molecules in the gas phase, Tech. rep., National
  Standard Reference Data System-National Bureau of Standards

\bibitem[{Nguyen {et~al.}(2014)Nguyen, Nguyen, Nguyen, Van~Hoang, \&
  Vereecken}]{nguyen2014theoretical}
Nguyen, H. M.~T., Nguyen, H.~T., Nguyen, T.-N., Van~Hoang, H., \& Vereecken, L.
  2014, JPCA, 118, 8861

\bibitem[{Niu {et~al.}(1993)Niu, Shirley, \& Bai}]{niu1993high}
Niu, B., Shirley, D.~A., \& Bai, Y. 1993, \jcp, 98, 4377

\bibitem[{Nixon {et~al.}(2010)Nixon, Achterberg, Teanby, Irwin, Flaud, Kleiner,
  Dehayem-Kamadjeu, Brown, Sams, B{\'e}zard, {et~al.}}]{nixon2010upper}
Nixon, C.~A., Achterberg, R.~K., Teanby, N.~A., {et~al.} 2010, Far. Dis., 147,
  65

\bibitem[{Oliveira {et~al.}(2016)Oliveira, Lehman, McCoy, \&
  Lineberger}]{oliveira2016photoelectron}
Oliveira, A.~M., Lehman, J.~H., McCoy, A.~B., \& Lineberger, W.~C. 2016, \jcp,
  145, 124317

\bibitem[{Olney {et~al.}(1997)Olney, Cann, Cooper, \&
  Brion}]{olney1997absolute}
Olney, T.~N., Cann, N., Cooper, G., \& Brion, C. 1997, Chem. Phys., 223, 59

\bibitem[{Phelps \& Dalby(1966)}]{phelps1966experimental}
Phelps, D., \& Dalby, F. 1966, \prl, 16, 3

\bibitem[{Phillips(1992)}]{phillips1992rate}
Phillips, L.~F. 1992, PrECS, 18, 75

\bibitem[{Potapov {et~al.}(2017)Potapov, Canosa, Jim{\'e}nez, \&
  Rowe}]{potapov2017uniform}
Potapov, A., Canosa, A., Jim{\'e}nez, E., \& Rowe, B. 2017, Angew.e Chem. Int.
  Ed., 56, 8618

\bibitem[{Quack \& Troe(1974)}]{quack1974specific}
Quack, M., \& Troe, J. 1974, Ber. Bunsengesellschaft Phys. Chem., 78, 240

\bibitem[{{Ruiterkamp} {et~al.}(2007){Ruiterkamp}, {Charnley}, {Butner},
  {Huang}, {Rodgers}, {Kuan}, \& {Ehrenfreund}}]{rui07}
{Ruiterkamp}, R., {Charnley}, S.~B., {Butner}, H.~M., {et~al.} 2007, \apss,
  310, 181

\bibitem[{{Rydbeck} {et~al.}(1973){Rydbeck}, {Elld{\'e}r}, \& {Irvine}}]{ryd73}
{Rydbeck}, O.~E.~H., {Elld{\'e}r}, J., \& {Irvine}, W.~M. 1973, \nat, 246, 466

\bibitem[{{Sandell} {et~al.}(1988){Sandell}, {Magnani}, \& {Lada}}]{sandell88}
{Sandell}, G., {Magnani}, L., \& {Lada}, E.~A. 1988, \apj, 329, 920

\bibitem[{Saxena {et~al.}(2003)Saxena, Bhatnagar, \&
  Singh}]{saxena2003production}
Saxena, P., Bhatnagar, S., \& Singh, M. 2003, Bull. Astron. Soc. India, 31, 67

\bibitem[{Shannon {et~al.}(2014)Shannon, Caravan, Blitz, \&
  Heard}]{shannon2014combined}
Shannon, R., Caravan, R., Blitz, M., \& Heard, D. 2014, PCCP, 16, 3466

\bibitem[{Shannon {et~al.}(2010)Shannon, Taylor, Goddard, Blitz, \&
  Heard}]{shannon2010observation}
Shannon, R.~J., Taylor, S., Goddard, A., Blitz, M.~A., \& Heard, D.~E. 2010,
  PCCP, 12, 13511

\bibitem[{Simnikov(1941)}]{Simnikov1941Increase}
Simnikov, I. 1941, Zh. Obshch. Khim., 14, 483

\bibitem[{Sivakumaran {et~al.}(2003)Sivakumaran, H{\"o}lscher, Dillon, \&
  Crowley}]{sivakumaran2003reaction}
Sivakumaran, V., H{\"o}lscher, D., Dillon, T.~J., \& Crowley, J.~N. 2003, PCCP,
  5, 4821

\bibitem[{Smith {et~al.}(2006{\natexlab{a}})Smith, Pope, Cronin, Parkes, \&
  Orr-Ewing}]{smith2006absorption}
Smith, C.~A., Pope, F.~D., Cronin, B., Parkes, C.~B., \& Orr-Ewing, A.~J.
  2006{\natexlab{a}}, JCPA, 110, 11645

\bibitem[{Smith(1980)}]{smith1980kinetics}
Smith, I.~W. 1980, Kinetics and Dynamics of Elementary Gas Reactions:
  Butterworths Monographs in Chemistry and Chemical Engineering (Butterworth
  and Co)

\bibitem[{Smith {et~al.}(2006{\natexlab{b}})Smith, Sage, Donahue, Herbst, \&
  Quan}]{smith2006temperature}
Smith, I.~W., Sage, A.~M., Donahue, N.~M., Herbst, E., \& Quan, D.
  2006{\natexlab{b}}, Far. Dis., 133, 137

\bibitem[{Snyder {et~al.}(1969)Snyder, Buhl, Zuckerman, \&
  Palmer}]{snyder1969microwave}
Snyder, L.~E., Buhl, D., Zuckerman, B., \& Palmer, P. 1969, \prl, 22, 679

\bibitem[{Stoecklin \& Clary(1995)}]{stoecklin1995fast}
Stoecklin, T., \& Clary, D. 1995, Journal of Molecular Structure: THEOCHEM,
  341, 53

\bibitem[{Stoecklin {et~al.}(1991)Stoecklin, Dateo, \&
  Clary}]{stoecklin1991rate}
Stoecklin, T., Dateo, C., \& Clary, D. 1991, Far. Trans., 87, 1667

\bibitem[{Taylor {et~al.}(2008)Taylor, Goddard, Blitz, Cleary, \&
  Heard}]{taylor2008pulsed}
Taylor, S.~E., Goddard, A., Blitz, M.~A., Cleary, P.~A., \& Heard, D.~E. 2008,
  PCCP, 10, 422

\bibitem[{Troe(1985)}]{troe1985statistical}
Troe, J. 1985, CPL, 122, 425

\bibitem[{{Tsang} \& {Hampson}(1986)}]{tsa86}
{Tsang}, W., \& {Hampson}, R.~F. 1986, JPCRD, 15, 1087

\bibitem[{{Turner}(1977)}]{tur77}
{Turner}, B.~E. 1977, \apj, 213, L75

\bibitem[{{Van de Sande} {et~al.}(2018){Van de Sande}, {Sundqvist}, {Millar},
  {Keller}, {Homan}, {de Koter}, {Decin}, \& {De Ceuster}}]{vds18a}
{Van de Sande}, M., {Sundqvist}, J.~O., {Millar}, T.~J., {et~al.} 2018, \aap,
  616, A106

\bibitem[{Viskari {et~al.}(2000)Viskari, Vartiainen, \&
  Pasanen}]{viskari2000seasonal}
Viskari, E.-L., Vartiainen, M., \& Pasanen, P. 2000, Atmos. Env., 34, 917

\bibitem[{{Wakelam} {et~al.}(2015){Wakelam}, {Loison}, {Herbst}, {Pavone},
  {Bergeat}, {B{\'e}roff}, {Chabot}, {Faure}, {Galli}, \& {Geppert}}]{wak15}
{Wakelam}, V., {Loison}, J.~C., {Herbst}, E., {et~al.} 2015, \apjs, 217, 20

\bibitem[{{Westley}(1980)}]{wes80}
{Westley}, F. 1980, {Table of Recommended Rate Constants for Chemical Reactions
  Occuring in Combustion }, Tech. rep., National Standard Reference Data
  System-National Bureau of Standards

\bibitem[{Zabarnick {et~al.}(1988)Zabarnick, Fleming, \&
  Lin}]{zabarnick1988temperature}
Zabarnick, S., Fleming, J., \& Lin, M. 1988, Symposium (International) on
  Combustion, 21, 713

\bibitem[{Zou {et~al.}(2004)Zou, Shu, Sears, Hall, \&
  North}]{zou2004photodissociation}
Zou, P., Shu, J., Sears, T.~J., Hall, G.~E., \& North, S.~W. 2004, The Journal
  of Physical Chemistry A, 108, 1482

\end{thebibliography}

%\listofchanges
\end{document}